\documentclass[journal]{IEEEtran}

\usepackage{textcomp}
\usepackage{xcolor}
\usepackage{graphicx}
\usepackage{amsmath,amssymb,amsfonts}
\usepackage{algorithm}
\usepackage{algorithmic}
\usepackage{cite}
\usepackage{lipsum}
\usepackage{epstopdf}
\usepackage{stfloats}
\usepackage{bm}
\usepackage{xcolor}
\usepackage{colortbl,booktabs}

\begin{document}
\title{Feasibility Study of UAV-Assisted Anti-Jamming Positioning}

\author{Zijie~Wang,
        Rongke~Liu,~\IEEEmembership{Senior Member,~IEEE,}
        Qirui~Liu,
        Lincong~Han
\thanks{This work was supported by the Beijing Municipal Science and Technology Project (Z181100003218008).}
\thanks{Z. Wang and R. Liu are with the School
of Electronic and Information Engineering, Beihang University, Beijing 100191,
China (e-mail: wangmajie@buaa.edu.cn; rongke\_liu@buaa.edu.cn).}
\thanks{Q. Liu is with the School
of Electronic and Information Engineering, Beihang University, Beijing 100191,
China.}
\thanks{L. Han is with the School
of Electronic and Information Engineering, Beihang University, Beijing 100191,
China.}
}

\markboth{Journal of \LaTeX\ Class Files,~Vol.~14, No.~8, August~2015}%
{Shell \MakeLowercase{\textit{et al.}}: Bare Demo of IEEEtran.cls for IEEE Journals}

\maketitle

\begin{abstract}
As the cost and technical difficulty of jamming devices continue to decrease, jamming has become one of the major threats to positioning service. Unfortunately, most conventional technologies are vulnerable to jamming attacks due to their inherent shortcomings like weak signal strength and unfavorable anchor geometry. Thanks to the high operational flexibility, unmanned aerial vehicle (UAV) could fly close to users to enhance signal strength while maintaining a satisfactory geometry, making it a potential solution to the above challenges. In this article, we propose a UAV-assisted anti-jamming positioning system, in which multiple UAVs first utilize time-difference-of-arrival (TDoA) measurements from ground reference stations and double-response two-way ranging (DR-TWR) measurements from UAV-to-UAV links to perform self-localization as well as clock synchronization, and then act as anchor nodes to provide TDoA positioning service for ground users in the presence of jamming. To evaluate the feasibility and performance of the proposed system, we first derive the Cram\'er-Rao lower bound (CRLB) of UAV self-localization. Then, the impacts of UAV position uncertainty and synchronization errors caused by jamming on positioning service are modeled, and the theoretical root-mean-square error (RMSE) of user position estimate is further derived. Numerical results demonstrate that the proposed system is a promising alternative to existing positioning systems when their services are disrupted by jamming. The most notable advantage of the proposed system is that it is fully compatible with existing user equipment (UE) and positioning methods.
\end{abstract}


\IEEEpeerreviewmaketitle

\section{Introduction}
\subsection{Motivation}
\IEEEPARstart{A}{s} people's demand for accurate location information continues to increase, positioning technologies are playing an increasingly important role in today's society \cite{LBS_Lives}. The use of positioning technologies enables a wide range of location-based services (LBS) like intelligent transport systems (ITS) and mobile marketing \cite{Pos_ITS}, thereby promoting the revolutionary development of industrial manufacturing and our daily lives. To this end, both the fifth generation (5G) wireless networks currently under construction and the future sixth generation (6G) networks have regarded positioning as a key technology and an essential service \cite{Pos_5G,Pos_6G}. Despite the above advantages, in practical applications, the availability and performance of positioning service can be severely affected by many factors, and jamming is one of them.

Generally speaking, jamming refers to a type of intentional radio frequency interference (RFI) emitted by hostile devices (also known as ``jammer''), whose aim is to block the positioning service or degrade accuracy by disturbing the signal reception at UE \cite{Pos_Jam}. Due to the rapid development of electronics industries in recent years, the cost and technical difficulty of jammer continue to decrease, a portable jammer could be easily obtained over the internet for less than 100 dollars \cite{Jam_Cheap_1}. Over the past five years, hundreds of jamming incidents have been reported around the world \cite{Jam_Report}. The frequent occurrence of jamming incidents poses serious threats to life-critical applications that require location information such as traffic management and emergency services.

Unfortunately, conventional wireless positioning technologies represented by global navigation satellite system (GNSS) and terrestrial cellular-based positioning are vulnerable to jamming attacks due to some inherent weakness. For the widely used GNSS systems, their positioning services rely on satellites operating in orbits of altitude ranging from 19100 to 35786km \cite{GNSS_Weak}. The long-distance propagation makes the strength of GNSS signals received at earth's surface extremely weak (only about -133 to -122dBm), which could be easily overwhelmed by jamming signals. In practical applications, a low-cost jammer with transmit power of 10dBm can disrupt all GNSS services within a radius of 100m \cite{Jam_Cheap_2}. Although many existing studies showed that the performance of GNSS receivers in jamming environments could be improved by adopting novel antennas or high-performance filters \cite{Anti_Antenna,Anti_Filter}, these approaches do not change the fact of weak signal strength, resulting in limited improvement in anti-jamming ability. Moreover, these approaches commonly require changes to receiver's hardware or software, which means that they are cost-consuming and incompatible with existing equipment. Similar to GNSS systems, terrestrial cellular-based positioning also has its own limitations. Since the cellular networks are originally designed for communication applications that only require the connection with one base station (BS), it is very difficult for users to find a sufficient number of BSs for positioning \cite{LTE_Num}. Even if the user could receive signals from multiple BSs, some of them may be far away from the user, resulting in weak signals that are vulnerable to jamming attacks \cite{LTE_Jam}. In addition, the geometry of terrestrial BSs is commonly unsuitable for positioning \cite{LTE_Geometry}, which leads to large position errors even under normal conditions. Thus, it is unrealistic to expect the terrestrial cellular-based positioning to perform well in the presence of jamming.

Due to the high operational flexibility and fully-controllable mobility, low-altitude unmanned aerial vehicle (UAV) has recently attracted increasing attention from the research community and is expected to bring a new paradigm for the design of wireless networks \cite{UAV_Paradigm}. In the field of communication, UAVs have been widely used as aerial BSs and relays to serve the ground users or coordinate with terrestrial networks. Except being used for communication, UAVs are also suitable for being employed as aerial anchor nodes to provide positioning services \cite{UAV_Anchor}, especially in jamming environments. Compared with satellites and BSs, UAVs are capable of flying close to users, so that enhance the received signal strength \cite{UAV_LoS}. Besides, through the optimization of UAV deployment strategy, users could easily establish connections with multiple UAVs, whose geometry could flexibility adjusted according to user's requirements. Therefore, it is very promising to utilize low-altitude UAVs and existing terrestrial infrastructures to develop a novel anti-jamming positioning system. The aim of this study is to design such a system and evaluate its feasibility.

\subsection{Related Work}
Due to the aforementioned advantages, UAV-enabled positioning has become a research hotspot in recent years. The concept of using UAVs as mobile anchor nodes to locate ground users and the corresponding positioning methods were introduced in \cite{UAV_Concept_1,UAV_Concept_2}. In \cite{UAV_HAWK}, \cite{UAV_GuideLoc}, two prototypes of UAV-enabled positioning system called HAWK and GuideLoc were proposed, which employ range-free approaches to obtain rough estimates of user's location. \emph{Sallouha et al}. \cite{UAV_RSS_1,UAV_RSS_2} and \emph{Wang et al}. \cite{UAV_TDoA} applied the received signal strength (RSS)-based and TDoA approaches to UAV-enabled positioning to improve the accuracy of location estimation. Furthermore, the service reliability of UAV-enabled positioning in mountainous environments are analyzed and enhanced in \cite{UAV_Reliable}. However, none of the above studies considered the impacts of jamming on positioning services, making these systems unreliable under jamming attacks. Moreover, most existing systems rely on GNSS systems to obtain the UAVs' locations, which is unrealistic in jamming environments \cite{Jam_Cheap_2}. In \cite{UAV_GRS}, ground reference stations (GRS) with known positions were used to locate UAVs, which is a potential solution to the problem of UAV self-localization in GNSS-denied environments. Nevertheless, jamming was still not taken into account in this research.

In most existing research on UAV-enabled positioning, the UAV's location information is commonly assumed to be perfectly known \cite{UAV_RSS_1,UAV_TDoA}. In practice, the self-localization and clock synchronization of UAVs rely on measurements provided by satellites, GRSs or other UAVs \cite{UAV_GRS,UAV_CLK}, which will also be inaccurate under jamming attacks. Therefore, in addition to affecting the positioning services for users, jamming will also cause anchor position and clock uncertainty in UAV-enabled positioning systems. The impacts of anchor position uncertainty on RSS-based positioning systems were analyzed with CRLB in \cite{APU_RSS}. In \cite{APU_ToA} and \cite{APU_TDoA_2}, the CRLBs of time-of-arrival (ToA) and TDoA positioning in the presence of anchor position uncertainty were derived. The CRLB used in these studies can be achieved by the well-known maximum-likelihood (ML) method, which is time-consuming and compute-intensive. The MSE of iterative least-squares (ILS) method derived in \cite{APU_TDoA_1} seems to be a better metric for evaluating the position accuracy of users with low-cost equipment. However, in \cite{APU_TDoA_1}, anchor nodes were assumed to be perfectly synchronized, which is a bit unreasonable in multi-UAV systems. Thus, to objectively evaluate the performance of a UAV-enabled positioning system in jamming environments, both the anchor position uncertainty and clock uncertainty should be taken into consideration.

\subsection{Main Contributions}
In this article, we propose a UAV-enabled anti-jamming positioning system consisting of multiple low-altitude UAV platform and GRSs to provide positioning services for users in jamming environments, and theoretically analyze its performance. The proposed system takes full advantages of UAV's high mobility and flexible deployment to improve the anti-jamming performance of positioning services without using complex algorithms or changing hardware, making it fully compatible with existing low-cost equipment. Specifically, the main contributions of this article are summarized as follows.
\begin{itemize}
\item We establish a very practical scenario in which both the UAV and ground user are affected by jamming. Compared with previous research, this scenario is more suitable for evaluating the anti-jamming performance of UAV-enabled positioning systems.
\item We study the problem of UAV self-localization and clock synchronization, which has been neglected by many existing research on UAV-enabled positioning. A hybrid TDoA/DR-TWR scheme is proposed to solve this problem, and the CRLB of UAV self-localization in the presence of jamming is derived.
\item We use the TDoA measurements provided by UAVs and ILS method to locate users. The impacts of UAV position and clock uncertainty on positioning service are analyzed, and the theoretical RMSE of user position estimate is derived. Compared with metrics used in existing research, the derived RMSE is more appropriate for describing the position accuracy of low-cost UE in UAV-enabled positioning systems.
\item We test the proposed system with several simulation experiments, and analyze the key factors affecting its anti-jamming performance. In particular, we give the UAVs' energy-efficient transmit power corresponding to a certain transmit power of GRSs.
\end{itemize}

The remainder of this article is organized as follows. The structure of the proposed system and the positioning methods used are given in Section II. Section III derives the CRLB of UAV self-localization and the RMSE of UE position estimate in jamming environments. Section IV provides numerical results to demonstrate the feasibility and validity of our proposed system. Finally, Section V concludes this article.

The main notations used in this article are summarized as follows. Scalars are denoted by italic letters ($a$). Column vectors and matrices are denoted by lowercase and uppercase boldface letters (${\bf{a}}$ and ${\bf{A}}$), respectively. The superscript $T$ indicates the transpose operation (${{\bf{A}}^T}$) and superscript $ - 1$ indicates matrix inverse (${{\bf{A}}^{ - 1}}$). $\left\|  \cdot  \right\|$ represents the Euclidean norm of a vector.

\section{System Design}
\begin{figure}[!t]
\centering
\includegraphics[height=1.95in,width=3.25in]{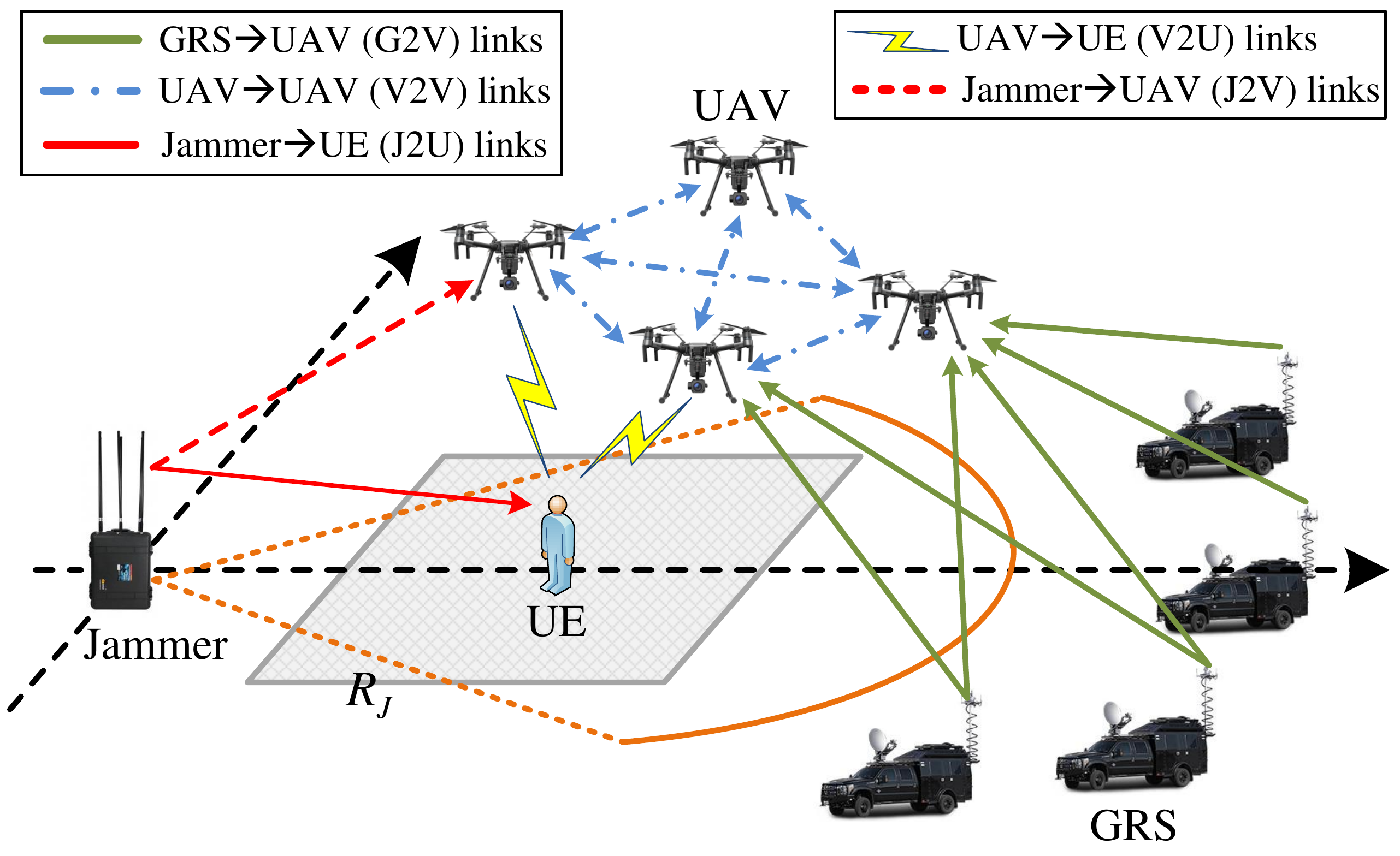}
\caption{Proposed UAV-assisted anti-jamming positioning system.}
\label{fig_1}
\end{figure}
In this article, as shown in Fig. 1, we consider a scenario consisting of a portable jammer, $M$ mobile GRSs and $N$ low-altitude UAV platforms. GRSs and UAVs are denoted by sets ${\cal G} \buildrel \Delta \over = \left\{ {{G_1},{G_2}, \cdots ,{G_M}} \right\}$ and ${\cal V} \buildrel \Delta \over = \left\{ {{V_1},{V_2}, \cdots ,{V_N}} \right\}$, respectively. The location of the jammer has been accurately measured in advance and is denoted by the horizontal coordinate ${\bf{w}} = {\left( {{x_J},{y_J}} \right)^T} \in {\mathbb{R}^{2 \times 1}}$ and the height ${h_J}$. Its three-dimensional (3-D) location is denoted by ${{\bf{w}}_{3D}} = {\left[ {{{\bf{w}}^T},{h_J}} \right]^T}$. The jammer continuously emits noise-like jamming signals in GNSS and 2.4 GHz ISM bands, blocking the reception of all GNSS signals within ${R_J}$ meters around it. This circular area in which GNSS services are completely disrupted is called ``jamming area'', whose boundary is indicated by the brown solid line in Fig. 1. In addition, the transmit power of the jammer in the ISM band is denoted by $P_J^t$.

A mobile GRS could be an autonomous land vehicle (ALV) equipped with a high-end GNSS receiver as well as an ISM band transceiver. The former is used to determine the GRS's own location, while the latter is used for providing positioning services. In order to receive GNSS signals and locate itself, each GRS must stay at least ${R_J}$ meters away from the jammer, that is, outside the jamming area. Moreover, we assume that if this requirement is satisfied, the impacts of jamming on GRSs' position accuracy can be effectively mitigated. The location of the \textit{m}-th GRS (${G_m}$) is denoted by the horizontal coordinate ${{\bf{g}}_m} = {\left( {x_G^m,y_G^m} \right)^T} \in {\mathbb{R}^{2 \times 1}}$ and the height ${h_G}$ (${{\bf{g}}_{m,3D}} = {\left[ {{\bf{g}}_m^T,{h_G}} \right]^T}$). $P_G^t$ is the transmit power of its ISM band transceiver.

The UE who needs positioning services stays in a square area called the ``target area'', which is marked in grey in Fig. 1. The center and side length of the target area are denoted by ${{\bf{o}}_T}$ and ${L_T}$, respectively. The prior probability that UE is located in each place within the target area is equal. The true location of the UE is denoted by ${{\bf{u}}^ * } \!=\! {\left( {x_U^ * ,y_U^ * } \right)^T} \!\in\! {\mathbb{R}^{2 \times 1}}$, and its height ${h_U}$ is set to 1.5m (${\bf{u}}_{3D}^ *  = {\left[ {{{\left( {{{\bf{u}}^ * }} \right)}^T},{h_U}} \right]^T}$), which is the average height of handheld devices. As can be seen from Fig. 1, the target area is inside the jamming area, which means that the UE cannot use GNSS systems to locate itself. As mentioned above, GRSs have to stay outside the jamming area, which makes them far away from the UE. Thus, the probability that LoS paths exist between GRSs and UE is extremely low. Moreover, in order to maintain the connection with the controller, GRSs are commonly not far apart, resulting in an unfavorable geometry for positioning. From the above analysis, it can be concluded that GRSs are also unsuitable for providing positioning services for UE in the target area.

In order to meet the UE's requirements for positioning service, we introduce UAVs into this scenario and form a novel anti-jamming positioning system together with existing GRSs. Similar to the UE, UAVs hovering at a fixed altitude ${h_V}$ over the jamming area cannot use GNSS systems to determine their own locations. Thus, each UAV is equipped with an ISM band transceiver, which will be used to establish wireless links with GRSs, other UAVs and the UE for self-localization as well as providing positioning services. The true location of the \textit{n}-th UAV (${V_n}$) is denoted by the horizontal coordinate ${\bf{v}}_n^ *  \!=\! {\left( {x_V^{n * },y_V^{n * }} \right)^T} \!\in\! {\mathbb{R}^{2 \times 1}}$ and the height ${h_V}$ (${\bf{v}}_{n,3D}^ *  \!=\! {\left[ {{{\left( {{\bf{v}}_n^ * } \right)}^T},{h_V}} \right]^T}$). The transmit power of its ISM band transceiver is denoted by $P_V^t$. In the proposed system, UAVs first utilize the measurements obtained from GRS-to-UAV (G2V) and UAV-to-UAV (V2V) links to perform self-localization and clock synchronization. After their own locations are determined, UAVs will be used as anchor nodes to provide positioning services for the UE through UAV-to-UE (V2U) links. Both the UAV self-localization process and service process are affected by jamming. Since UAVs could get close to the UE and maintain a satisfactory geometry, the proposed system is expected to achieve good anti-jamming performance.

\subsection{Hybrid TDoA/DR-TWR UAV Self-Localization}
As shown in Fig. 1, in the proposed system, each UAV ${V_n}$ could use its ISM band transceiver to establish $M$ G2V measurement links and $N - 1$ V2V measurement links. Moreover, there is also a Jammer-to-UAV (J2V) jamming link between the jammer and each UAV. These three kinds of wireless links can be characterized with two types of channels, namely the Ground-to-Air (G2A) channel and Air-to-Air (A2A) channel. The former includes the G2V and J2V links, while the V2V links belong to the latter. The G2V links are assumed to be dominated by LoS components. This assumption is quite reasonable because the high altitude of UAVs commonly leads to a high probability of LoS propagation \cite{G2A_Channel}. Then, the average path loss between GRS ${G_m}$ and UAV ${V_n}$ can be expressed as
\begin{equation}
P{L_{{G_m} \to {V_n}}} = {\beta _0}{\left( {\left\| {{\bf{v}}_{n,3D}^ *  - {{\bf{g}}_{m,3D}}} \right\|} \right)^{\alpha _{G2A}^L}},
\end{equation}
where ${\beta _0} = {\left( {{\textstyle{{4\pi {f_c}} \over c}}} \right)^2}$ is the reference path loss at a distance of 1m; ${f_c}$ and $c$ are the main frequency of the ISM band transceiver (2.4GHz) and the speed of light, respectively. $\alpha _{G2A}^L$ is the path loss exponent (PLE) of the G2A channel under LoS conditions.

Moreover, we also assume that there are always clear LoS paths between UAVs \cite{A2A_Channel}. Therefore, the path loss between UAV ${V_n}$ and ${V_i}$ follows the free space propagation model and can be written as
\begin{equation}
P{L_{{V_i} \to {V_n}}} = {\beta _0}{\left( {\left\| {{\bf{v}}_{n.3D}^ *  - {\bf{v}}_{i,3D}^ * } \right\|} \right)^2}.
\end{equation}

Different from G2V links, the propagation condition of the J2V link could be LoS or NLoS. The NLoS condition occurs when the UAV chooses to hide behind obstructions like buildings or mountains to reduce the impact of jamming on its signal reception. Then, the average path loss between the jammer and UAV ${V_n}$ can be expressed as
\begin{equation}
P{L_{J \to {V_n}}} = {\beta _0}{\left( {\left\| {{\bf{v}}_{n,3D}^ *  - {{\bf{w}}_{3D}}} \right\|} \right)^{\alpha _{G2A}^X}},
\end{equation}
where the superscript $X$ of $\alpha _{G2A}^X$ is either L or N, indicating the propagation condition (LoS or NLoS).

The signal-to-interference-plus-noise ratios (SINR) of the positioning signals transmitted by GRS ${G_m}$ and UAV ${V_i}$ at UAV ${V_n}$ can be expressed as follows:
\begin{equation}
SIN{R_{{G_m} \to {V_n}}} = \frac{{{{P_G^t} \mathord{\left/
 {\vphantom {{P_G^t} {P{L_{{G_m} \to {V_n}}}}}} \right.
 \kern-\nulldelimiterspace} {P{L_{{G_m} \to {V_n}}}}}}}{{{P_{{n_0}}} + {{P_J^t} \mathord{\left/
 {\vphantom {{P_J^t} {P{L_{J \to {V_n}}}}}} \right.
 \kern-\nulldelimiterspace} {P{L_{J \to {V_n}}}}}}},
\end{equation}
\begin{equation}
SIN{R_{{V_i} \to {V_n}}} = \frac{{{{P_V^t} \mathord{\left/
 {\vphantom {{P_V^t} {P{L_{{V_i} \to {V_n}}}}}} \right.
 \kern-\nulldelimiterspace} {P{L_{{V_i} \to {V_n}}}}}}}{{{P_{{n_0}}} + {{P_J^t} \mathord{\left/
 {\vphantom {{P_J^t} {P{L_{J \to {V_n}}}}}} \right.
 \kern-\nulldelimiterspace} {P{L_{J \to {V_n}}}}}}}.
\end{equation}
where ${P_{{n_0}}}$ is the noise power. As described in \cite{ToA_Var}, the minimum variances of ToA measurements that UAV ${V_n}$ could achieve are given by
\begin{figure}[!t]
\centering
\includegraphics[height=1.80in,width=3.05in]{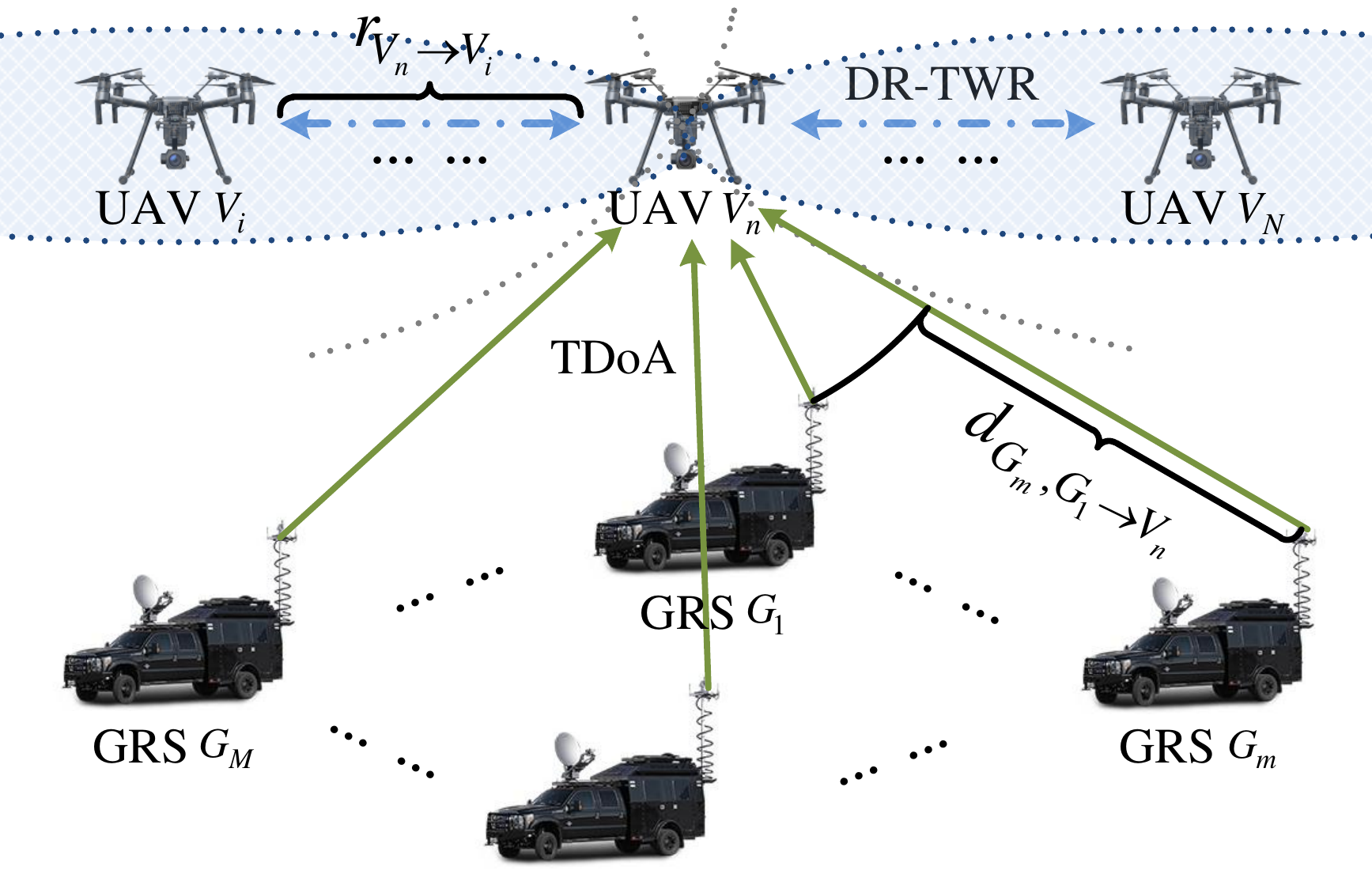}
\caption{Model of UAV self-localization.}
\label{fig_2}
\end{figure}
\begin{equation}
\sigma _{{G_m} \to {V_n}}^2\left( {{m^2}} \right) = {{{c^2}} \mathord{\left/
 {\vphantom {{{c^2}} {\left( {{B^2} \cdot SIN{R_{{G_m} \to {V_n}}}} \right)}}} \right.
 \kern-\nulldelimiterspace} {\left( {{B^2} \cdot SIN{R_{{G_m} \to {V_n}}}} \right)}},
\end{equation}
\begin{equation}
\sigma _{{V_i} \to {V_n}}^2\left( {{m^2}} \right) = {{{c^2}} \mathord{\left/
 {\vphantom {{{c^2}} {\left( {{B^2} \cdot SIN{R_{{V_i} \to {V_n}}}} \right)}}} \right.
 \kern-\nulldelimiterspace} {\left( {{B^2} \cdot SIN{R_{{V_i} \to {V_n}}}} \right)}},
\end{equation}
where $B$ is the signal bandwidth. As shown in Fig. 2, UAVs utilize two types of measurements to estimate their own locations, one of which is the TDoA measurement obtained through G2V links. Since GRSs have been accurately synchronized with each other using GNSS systems, measuring the TDoA between a pair of GRSs could eliminate the unknown clock bias between GRSs and the UAV. Let GRS ${G_1}$ be the reference node, the TDoA measurement of GRS pair $\left\langle {{G_m},{G_1}} \right\rangle $ measured at UAV ${V_n}$ can be expressed as
\begin{equation}
\begin{split}
{d_{{G_{m,}}\!{G_1} \!\to\! {V_n}}} &=\! \left\|\! {{\bf{v}}_{n\!,\!3D}^ *  \!-\! {{\bf{g}}_{m\!,\!3D}}} \!\right\| \!-\! \left\|\! {{\bf{v}}_{n\!,\!3D}^ *  \!-\! {{\bf{g}}_{1\!,\!3D}}} \!\right\| \!+\! {n_{{G_m}\!,\!{G_1} \!\to\! {V_n}}}\\
&= r_{{G_m} \to {V_n}}^ *  - r_{{G_1} \to {V_n}}^ *  + {n_{{G_m},{G_1} \to {V_n}}}\\
&= d_{{G_{m,}}{G_1} \to {V_n}}^ *  + {n_{{G_m},{G_1} \to {V_n}}},
\end{split}
\end{equation}
where $d_{{G_{m,}}{G_1} \to {V_n}}^ * $ denotes the true TDoA of GRS pair $\left\langle {{G_m},{G_1}} \right\rangle $; $r_{{G_m} \to {V_n}}^ * $ ($r_{{G_1} \to {V_n}}^ * $) is the true distance between GRS ${G_m}$ (${G_1}$) and UAV ${V_n}$; ${n_{{G_m},{G_1} \to {V_n}}} \sim {\cal N}\left( {0,\sigma _{{G_m} \to {V_n}}^2 + \sigma _{{G_1} \to {V_n}}^2} \right)$ is the TDoA measurement error caused by transceiver's internal noise and jamming.

Through G2V links, each UAV ${V_n}$ could collect $M - 1$ TDoA measurements, which can be represented by the following vector:
\begin{equation}
\begin{split}
{{\bf{d}}_{G \to {V_n}}} &= {\left[ {{d_{{G_2},{G_1} \to {V_n}}}, \cdots ,{d_{{G_M},{G_1} \to {V_n}}}} \right]^T}\\
&= {\bf{d}}_{G \to {V_n}}^ *  + {{\bf{n}}_{G \to {V_n}}},
\end{split}
\end{equation}
where vector ${\bf{d}}_{G \to {V_n}}^ *  = {\left[ {d_{{G_{2,}}{G_1} \to {V_n}}^ * , \cdots ,d_{{G_{M,}}{G_1} \to {V_n}}^ * } \right]^T}$ denotes the true values of $M - 1$ TDoA measurements, and ${{\bf{n}}_{G \to {V_n}}} = {\left[ {{n_{{G_2},{G_1} \to {V_n}}}, \cdots ,{n_{{G_M},{G_1} \to {V_n}}}} \right]^T}$ is the vector of measurement errors. The total TDoA measurements collected by $N$ UAVs form the following $N\left( {M - 1} \right) \times 1$ vector:
\begin{equation}
{{\bf{d}}_{G \to V}} = {\left[ {{\bf{d}}_{G \to {V_1}}^T, \cdots ,{\bf{d}}_{G \to {V_N}}^T} \right]^T} = {\bf{d}}_{G \to V}^ *  + {{\bf{n}}_{G \to V}},
\end{equation}
where ${\bf{d}}_{G \to V}^ *  \!=\! {\left[ {{{\left( {{\bf{d}}_{G \to {V_1}}^ * } \right)}^T}\!,\! \cdots ,{{\left( {{\bf{d}}_{G \to {V_N}}^ * } \right)}^T}} \right]^T}$ and ${{\bf{n}}_{G \to V}} = {\left[ {{\bf{n}}_{G \!\to\! {V_1}}^T, \cdots ,{\bf{n}}_{G \to {V_N}}^T} \right]^T}$.

The other type of measurements for UAV self-localization is the range measurement obtained with DR-TWR technique through V2V links. DR-TWR technique can ease the constraint of clock synchronization through exchange of messages \cite{DR_TWR}, making it suitable for measuring the range between two UAVs before they are synchronized. As will be derived in Appendix A, the DR-TWR measurement corresponding to UAV pair $\left\langle {{V_n},{V_i}} \right\rangle $ ($i \ne n$) and obtained at UAV ${V_n}$ can be expressed as
\begin{equation}
{r_{{V_n} \to {V_i}}} = r_{{V_n} \to {V_i}}^ *  \!+\! {n_{{V_n} \to {V_i}}} = \left\| {{\bf{v}}_n^ *  \!-\! {\bf{v}}_i^ * } \right\| \!+\! {n_{{V_n} \to {V_i}}},
\end{equation}
where $r_{{V_n} \to {V_i}}^ * $ denotes the true range between UAV ${V_n}$ and ${V_i}$; ${n_{{V_n} \!\to\! {V_i}}} \!\sim\! {\cal N}\left( {0,\frac{1}{4}\sigma _{{V_n} \!\to\! {V_i}}^2 \!+\! \frac{5}{4}\sigma _{{V_i} \!\to\! {V_n}}^2} \right)$ is the range measurement error caused by transceivers' internal noise and jamming.

Let ${{\cal L}_n} \buildrel \Delta \over = \left\{ {1, \cdots ,n - 1,n + 1, \cdots ,N} \right\}$, and ${L_{n,i}}$ represents the \textit{i}-th element in set ${{\cal L}_n}$. Then, the $N - 1$ DR-TWR range measurements collected by UAV ${V_n}$ can be represented by the vector:
\begin{equation}
\begin{split}
{{\bf{r}}_{{V_n} \to V}} &= {\left[ {{r_{{V_n} \to {V_{{L_{n,1}}}}}}, \cdots ,{r_{{V_n} \to {V_{{L_{n,N - 1}}}}}}} \right]^T}\\
&= {\bf{r}}_{{V_n} \to V}^ *  + {{\bf{n}}_{{V_n} \to V}},
\end{split}
\end{equation}
where vector ${\bf{r}}_{{V_n} \!\to\! V}^ *  \!=\! {\left[\! {r_{{V_n} \!\to\! {V_{{L_{n,1}}}}}^ * \!,\! \cdots \!,\!r_{{V_n} \!\to\! {V_{{L_{n,N - 1}}}}}^ * } \!\right]^T}$ denotes the true values of $N - 1$ range measurements, and ${{\bf{n}}_{{V_n} \!\to\! V}} \!=\! {\left[ {{n_{{V_n} \!\to\! {V_{{L_{n,1}}}}}}, \cdots ,{n_{{V_n} \!\to\! {V_{{L_{n,N - 1}}}}}}} \right]^T}$ is the vector of measurement errors. The total $N\left( {N - 1} \right)$ range measurements obtained by $N$ UAVs form the following vector:
\begin{equation}
{{\bf{r}}_{V \to V}} = {\left[ {{\bf{r}}_{{V_1} \to V}^T, \!\cdots\! ,{\bf{r}}_{{V_N} \to V}^T} \right]^T} \!=\! {\bf{r}}_{V \to V}^ *  + {{\bf{n}}_{V \to V}},
\end{equation}
\newcounter{mytempeqncnt}
\begin{figure*}[!b]
\vspace*{4pt}
\hrulefill
\normalsize
\setcounter{mytempeqncnt}{\value{equation}}
\setcounter{equation}{19}
\begin{equation}
\begin{split}
{d_{{V_n},{V_1} \to U}} &= \left\| {{\bf{u}}_{3D}^ *  - {\bf{v}}_{n,3D}^ * } \right\| - \left\| {{\bf{u}}_{3D}^ *  - {\bf{v}}_{1,3D}^ * } \right\| - \left( {\Delta {t _{{V_n}}} - \Delta {t _{{V_1}}}} \right) + {n_{{V_n},{V_1} \to U}}\\
&= d_{{V_n},{V_1} \to U}^ *  - \left[ {\left( {{{\hat r}_{{G_1} \to {V_n}}} - r_{{G_1} \to {V_n}}^ * } \right) - \left( {{{\hat r}_{{G_1} \to {V_1}}} - r_{{G_1} \to {V_1}}^ * } \right)} \right] - \left( {{e_{{G_1} \to {V_n}}} - {e_{{G_1} \to {V_1}}}} \right) + {n_{{V_n},{V_1} \to U}}\\
&= d_{{V_n},{V_1} \to U}^ *  - \Delta t _{{V_n},{V_1} \to U}^{Pos} - \Delta t _{{V_n},{V_1} \to U}^{Noi} + {n_{{V_n},{V_1} \to U}}.
\end{split}
\end{equation}
\setcounter{equation}{\value{mytempeqncnt}}
\end{figure*}

Putting the measurement vectors ${{\bf{d}}_{G \to V}}$ and ${{\bf{r}}_{V \to V}}$ together, the total $N\left( {M + N - 2} \right)$ measurements obtained by UAVs can be represented by the vector:
\begin{equation}
{{\bf{o}}_V} = {\left[ {{\bf{d}}_{G \to V}^T,{\bf{r}}_{V \to V}^T} \right]^T} = {\bf{o}}_V^ *  + {{\bf{n}}_V},
\end{equation}
where ${\bf{o}}_V^ *  \!=\! {\left[ {{{\left(\! {{\bf{d}}_{G \to V}^ * } \!\right)}^T},{{\left(\! {{\bf{r}}_{V \to V}^ * } \!\right)}^T}} \right]^T}$, ${{\bf{n}}_V} \!=\! {\left[ {{\bf{n}}_{G \to V}^T,{\bf{n}}_{V \to V}^T} \right]^T}$. The parameters to be estimated in the UAV self-localization process are the horizontal coordinates of $N$ UAVs, which can be denoted as
\begin{equation}
{\bf{v}} = {\left[ {{\bf{v}}_1^T, \cdots ,{\bf{v}}_N^T} \right]^T} = {\left[ {x_V^1,y_V^1, \cdots ,x_V^N,y_V^N} \right]^T}.
\end{equation}

In the proposed system, UAVs send all their measurements (${{\bf{o}}_V}$) to GRS ${G_1}$, which will use the well-known ML method to estimate UAVs' locations (${\bf{v}}$). The estimated locations are denoted by vector ${\bf{\hat v}} = {\left[ {{\bf{\hat v}}_1^T, \cdots ,{\bf{\hat v}}_N^T} \right]^T}$.

\subsection{Clock Synchronization between UAVs}
To be employed as anchor nodes for TDoA positioning, UAVs need to be clock synchronized. In the proposed system, the local clock of GRS ${G_1}$ that has already been synchronized with GNSS is treated as the reference clock for timing services. During the mission, GRS ${G_1}$ periodically sends synchronization messages with timestamps to UAVs. The synchronization message sent at time ${t_s}$ will be detected by UAV ${V_n}$ at time
\begin{equation}
t_{r,n}^ *  = {t_s} + {{\left( {r_{{G_1} \to {V_n}}^ *  + {e_{{G_1} \to {V_n}}}} \right)} \mathord{\left/
 {\vphantom {{\left( {r_{{G_1} \to {V_n}}^ *  + {e_{{G_1} \to {V_n}}}} \right)} c}} \right.
 \kern-\nulldelimiterspace} c},
\end{equation}
where ${e_{{G_1} \to {V_n}}} \sim {\cal N}\left( {0,\sigma _{{G_1} \to {V_n}}^2} \right)$ denotes the synchronization error (m) caused by transceiver's internal noise and jamming.

For UAV ${V_n}$, the transmission time ${t_s}$ of the synchronization message can be extracted from the received timestamp, while the true range $r_{{G_1} \to {V_n}}^ * $ is unavailable. Replace $r_{{G_1} \to {V_n}}^ * $ in equation (16) with the predicted range ${\hat r_{{G_1} \to {V_n}}}$ based on the estimated location ${{\bf{\hat v}}_{n,3D}} = {\left[ {{\bf{\hat v}}_n^T,{h_V}} \right]^T}$, the synchronization result at UAV ${V_n}$ can be expressed as
\begin{equation}
{\hat t_{r,{V_n}}} = {t_s} + \frac{{{{\hat r}_{{G_1} \to {V_n}}}}}{c} = {t_s} + \frac{{\left\| {{{{\bf{\hat v}}}_{n,3D}} - {{\bf{g}}_{1,3D}}} \right\|}}{c}.
\end{equation}
Then, the clock synchronization error (m) corresponding to UAV ${V_n}$ can be written as
\begin{equation}
\begin{split}
\Delta {t _{{V_n}}} &=\! c \!\cdot\! \left( {{{\hat t}_{r,{V_n}}} \!-\! t_{r,{V_n}}^ * } \right) \!=\! {{\hat r}_{{G_1} \!\to\! {V_n}}} \!-\! r_{{G_1} \!\to\! {V_n}}^ *  \!+\! {e_{{G_1} \!\to\! {V_n}}}\\
&=\! \left\|\! {{{{\bf{\hat v}}}_{n,3D}} \!-\! {{\bf{g}}_{1,3D}}} \!\right\| \!-\! \left\|\! {{\bf{v}}_{n,3D}^ *  - {{\bf{g}}_{1,3D}}} \!\right\| \!+\! {e_{{G_1} \to {V_n}}},
\end{split}
\end{equation}

It can be clearly seen from the above equation that the clock synchronization error consists of two components, that is, the range prediction error caused by UAV position uncertainty and the ToA measurement error caused by internal noise and jamming.

\subsection{TDoA Positioning Services for UE}
\begin{figure}[!t]
\centering
\includegraphics[height=1.60in,width=2.60in]{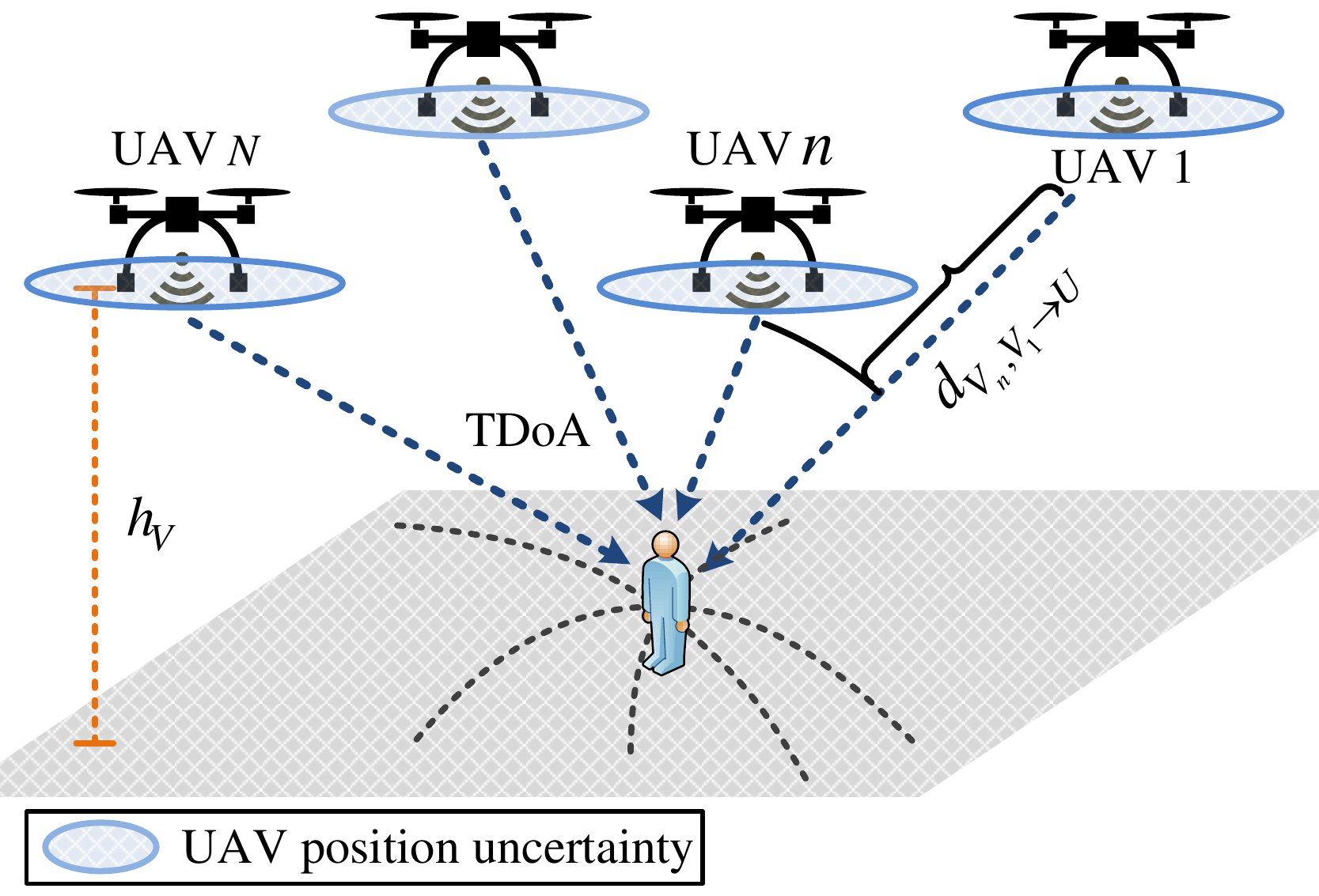}
\caption{Model of UE Positioning.}
\label{fig_3}
\end{figure}
After their time and locations are determined by the schemes introduced in the previous subsections, UAVs in the proposed system will be used as anchor nodes to provide positioning services for the UE. As shown in Fig. 1, the UE could receive positioning signals from UAVs through $N$ UAV-to-UE (V2U) measurement links. Similar to UAVs, the signal receptions at the UE are also affected by the Jammer-to-UE (J2U) jamming link. The V2U links can be modeled as G2A channels, and their propagation conditions are assumed to be LoS. Thus, the average path loss $P{L_{{V_n} \to U}}$ between the UE and UAV ${V_n}$ can be calculated by replacing ${{\bf{g}}_{m,3D}}$ in equation (1) with ${\bf{u}}_{3D}^ * $. The J2U jamming link is a typical Ground-to-Ground (G2G) channel. Since the jammer is close to the UE, the J2U link is assumed to be dominated by the LoS component, and its average path loss can be expressed as
\begin{equation}
P{L_{J \to U}} = {\beta _0}{\left( {\left\| {{\bf{u}}_{3D}^ *  - {{\bf{w}}_{3D}}} \right\|} \right)^{\alpha _{G2G}^L}},
\end{equation}
where $\alpha _{G2G}^L$ denotes the PLE of the G2G channel under LoS conditions. Then, the SINR ($SIN{R_{{V_n} \to U}}$) and ToA measurement variance ($\sigma _{{V_n} \to U}^2$) at the UE can be calculated with approaches similar to equations (4) and (6).

As shown in Fig. 3, we use the TDoA technique to support positioning services. The reason for choosing TDoA is that this technique has been widely adopted in many existing systems like LTE and NB-IoT networks, so that the services provided by the proposed system can be fully compatible with existing equipment. Let UAV ${V_1}$ be the reference node, the TDoA measurement of UAV pair $\left\langle {{V_n},{V_1}} \right\rangle $ measured at the UE can be written as equation (20), where $d_{{V_n},{V_1} \to U}^ * $ denotes the true value of TDoA measurement; $\Delta t _{{V_n},{V_1} \to U}^{Pos} \!=\! \left( {{{\hat r}_{{G_1} \!\to\! {V_n}}} \!-\! r_{{G_1} \!\to\! {V_n}}^ * } \right) \!-\! \left( {{{\hat r}_{{G_1} \!\to\! {V_1}}} \!-\! r_{{G_1} \to {V_1}}^ * } \right)$ and $\Delta t _{{V_n},{V_1} \to U}^{Noi} \!=\! \left( {{e_{{G_1} \!\to\! {V_n}}} \!-\! {e_{{G_1} \!\to\! {V_1}}}} \right) \!\sim\! {\cal N}\left( {0,\sigma _{{G_1} \!\to\! {V_n}}^2 \!+\! \sigma _{{G_1} \!\to\! {V_1}}^2} \right)$ indicates the impacts of clock synchronization errors caused by UAV position uncertainty and jamming on the TDoA measurement, respectively; ${n_{{V_n},{V_1} \!\to\! U}} \!\sim\! {\cal N}\left( {0,\sigma _{{V_n} \!\to\! U}^2 \!+\! \sigma _{{V_1} \!\to\! U}^2} \right)$ is the measurement error caused by UE's internal noise and jamming.

The UE could collect $N - 1$ TDoA measurements through V2U links, which can be represented by the following vector:
\begin{equation}\tag{21}
\begin{split}
{{\bf{d}}_{V \to U}} &= {\left[ {{d_{{V_2},{V_1} \to U}}, \cdots ,{d_{{V_N},{V_1} \to U}}} \right]^T}\\
&= {\bf{d}}_{V \to U}^ *  - \Delta {\bf{t }}_{V \to U}^{Pos} - \Delta {\bf{t }}_{V \to U}^{Noi} + {{\bf{n}}_{V \to U}},
\end{split}
\end{equation}
\newcounter{mytempeqncntt}
\begin{figure*}[!b]
\vspace*{4pt}
\hrulefill
\normalsize
\setcounter{mytempeqncntt}{\value{equation}}
\setcounter{equation}{37}
\begin{equation}
{{\bf{R}}_n}\left( {i,:} \right) \!=\! \frac{{\partial r_{{V_n} \to {V_{{L_{n,i}}}}}^ * }}{{\partial {\bf{v}}}} \!=\! \left\{ {\begin{array}{*{20}{c}}
{\left[ {{\bf{0}}_{2\left( {{L_{n,i}} - 1} \right) \times 1}^T, - {{\left(\! {{\bf{k}}_{{\bf{v}}_{{L_{n,i}}}^ * }^{{\bf{v}}_n^ * }} \!\right)}^T},{\bf{0}}_{2\left( {n - {L_{n,i}} - 1} \right) \times 1}^T,{{\left(\! {{\bf{k}}_{{\bf{v}}_{{L_{n,i}}}^ * }^{{\bf{v}}_n^ * }} \!\right)}^T},{\bf{0}}_{2\left( {N - n} \right) \times 1}^T} \right],\;\;{\rm{  if  }}\;\,{L_{n,i}} < n,}\\
{\left[ {{\bf{0}}_{2\left( {n - 1} \right) \times 1}^T,{{\left(\! {{\bf{k}}_{{\bf{v}}_{_{{L_{n,i}}}}^ * }^{{\bf{v}}_n^ * }} \!\right)}^T},{\bf{0}}_{2\left( {{L_{n,i}} - n - 1} \right) \times 1}^T, - {{\left(\! {{\bf{k}}_{{\bf{v}}_{_{{L_{n,i}}}}^ * }^{{\bf{v}}_n^ * }} \!\right)}^T},{\bf{0}}_{2\left( {N - {L_{n,i}}} \right) \times 1}^T} \right],\;\;{\rm{  if  }}\;\,{L_{n,i}} \ge n,}
\end{array}} \right.
\end{equation}
\setcounter{equation}{\value{mytempeqncntt}}
\end{figure*}
where
\begin{equation}\tag{22}
{\bf{d}}_{V \to U}^ *  = {\left[ {d_{{V_2},{V_1} \to U}^ * , \cdots ,d_{{V_N},{V_1} \to U}^ * } \right]^T},
\end{equation}
\begin{equation}\tag{23}
\Delta {\bf{t}}_{V \to U}^{Pos} = {\left[ {\Delta t_{{V_2},{V_1} \to U}^{Pos}, \cdots ,\Delta t_{{V_N},{V_1} \to U}^{Pos}} \right]^T},
\end{equation}
\begin{equation}\tag{24}
\Delta {\bf{t}}_{V \to U}^{Noi} = {\left[ {\Delta t_{{V_2},{V_1} \to U}^{Noi}, \cdots ,\Delta t_{{V_N},{V_1} \to U}^{Noi}} \right]^T},
\end{equation}
\begin{equation}\tag{25}
{{\bf{n}}_{V \to U}} = {\left[ {{n_{{V_2},{V_1} \to U}}, \cdots ,{n_{{V_N},{V_1} \to U}}} \right]^T}.
\end{equation}

After collecting all the TDoA measurements in vector ${{\bf{d}}_{V \to U}}$, the horizontal coordinate (${\bf{u}} = {\left[ {{x_U},{y_U}} \right]^T}$) of the UE could be estimated with the widely used ILS method. The estimated UE location is denoted by vector ${\bf{\hat u}} = {\left[ {{{\hat x}_U},{{\hat y}_U}} \right]^T}$.

\section{Analysis of Theoretical Performance}
In this section, we evaluate the theoretical performance of the proposed system under jamming attacks. Different from previous studies in which the UAVs' locations are assumed to be perfectly known, we consider the impacts of jamming on both UAV self-localization and UE position estimation, making the evaluation results convincing. Specifically, the CRLB of the proposed hybrid TDoA/DR-TWR UAV self-localization scheme is first derived in subsection A. Then, in subsection B, we derive the RMSE of UE position estimation in the presence of UAV position and clock uncertainty.

\subsection{CRLB of UAV Self-Localization}
As mentioned in Section II.A, during the UAV self-localization process, GRS ${G_1}$ with sufficient computer power will use the ML method to determine UAVs' locations. Thus, the CRLB that could be achieved by ML method is used to indicate the theoretical performance of UAV self-localization in the presence of jamming. Since the TDoA vector ${{\bf{d}}_{G \to V}}$ and DR-TWR vector ${{\bf{r}}_{V \to V}}$ are independent of each other, the log-likelihood function of the measurement vector ${{\bf{o}}_V}$ can be expressed as
\begin{equation}\tag{26}
\ln f\left( {{{\bf{o}}_V};{\bf{v}}} \right) = \ln f\left( {{{\bf{d}}_{G \to V}};{\bf{v}}} \right) + \ln f\left( {{{\bf{r}}_{V \to V}};{\bf{v}}} \right),
\end{equation}
where $f\left( {{{\bf{d}}_{G \to V}};{\bf{v}}} \right)$ and $f\left( {{{\bf{r}}_{V \to V}};{\bf{v}}} \right)$ are likelihood functions of TDoA measurements and DR-TWR measurements, respectively. The expressions of $f\left( {{{\bf{d}}_{G \to V}};{\bf{v}}} \right)$ and $f\left( {{{\bf{r}}_{V \to V}};{\bf{v}}} \right)$ are
\begin{equation}\tag{27}
\begin{split}
f\left( {{{\bf{d}}_{G \to V}};{\bf{v}}} \right) = {\left( {{{\left( {2\pi } \right)}^{N\left( {M - 1} \right)}}\left| {{{\bf{Q}}_{{{\bf{n}}_{G \to V}}}}} \right|} \right)^{\frac{1}{2}}}\qquad\qquad\;\;\;\\
\quad\cdot\! \exp \left(\! { \!-\! \frac{1}{2}{{\left(\! {{{\bf{d}}_{G \!\to\! V}} \!-\! {\bf{d}}_{G \!\to\! V}^ * } \!\right)}^T}{\bf{Q}}_{{{\bf{n}}_{G \!\to\! V}}}^ * \left(\! {{{\bf{d}}_{G \!\to\! V}} \!-\! {\bf{d}}_{G \!\to\! V}^ * } \!\right)} \!\right),
\end{split}
\end{equation}
\begin{equation}\tag{28}
\begin{split}
f\left( {{{\bf{r}}_{V \to V}};{\bf{v}}} \right) = {\left( {{{\left( {2\pi } \right)}^{N\left( {N - 1} \right)}}\left| {{{\bf{Q}}_{{{\bf{n}}_{V \to V}}}}} \right|} \right)^{\frac{1}{2}}}\qquad\qquad\;\;\;\\
\quad\cdot\! \exp \left(\! { \!-\! \frac{1}{2}{{\left(\! {{{\bf{r}}_{V \!\to\! V}} \!-\! {\bf{r}}_{V \!\to\! V}^ * } \!\right)}^T}{\bf{Q}}_{{{\bf{n}}_{V \!\to\! V}}}^ * \left(\! {{{\bf{r}}_{V \!\to\! V}} \!-\! {\bf{r}}_{V \!\to\! V}^ * } \!\right)} \!\right),
\end{split}
\end{equation}
where ${{\bf{Q}}_{{{\bf{n}}_{G \to V}}}}$ and ${{\bf{Q}}_{{{\bf{n}}_{V \to V}}}}$ are the covariance matrices of TDoA error vector ${{\bf{n}}_{G \to V}}$ and DR-TWR error vector ${{\bf{n}}_{V \to V}}$, respectively. Their expressions can be written as
\begin{equation}\tag{29}
{{\bf{Q}}_{{{\bf{n}}_{G \!\to\! V}}}} \!=\! {\mathop{\rm cov}} \left(\! {{{\bf{n}}_{G \!\to\! V}}} \!\right) \!=\! {\mathop{\rm blkdiag}\nolimits}\! \left(\! {{{\bf{Q}}_{{{\bf{n}}_{G \!\to\! {V_1}}}}}\!,\! \cdots \!,\!{{\bf{Q}}_{{{\bf{n}}_{G \!\to\! {V_N}}}}}} \!\right),
\end{equation}
\begin{equation}\tag{30}
{{\bf{Q}}_{{{\bf{n}}_{V \!\to\! V}}}} \!=\! {\mathop{\rm cov}} \left(\! {{{\bf{n}}_{V \!\to\! V}}} \!\right) \!=\! {\mathop{\rm blkdiag}\nolimits}\! \left(\! {{{\bf{Q}}_{{{\bf{n}}_{{V_1} \!\to\! V}}}}\!,\! \cdots \!,\!{{\bf{Q}}_{{{\bf{n}}_{{V_N} \!\to\! V}}}}} \!\right),
\end{equation}
where
\begin{equation}\tag{31}
{{\bf{Q}}_{{{\bf{n}}_{G \!\to\! {V_n}}}}} \!=\! \sigma _{{G_1} \to {V_n}}^2 \!\cdot\! {{\bf{I}}_{M \!-\! 1}} \!+\! {\mathop{\rm diag}\nolimits} \!\left( {\sigma _{{G_2} \!\to\! {V_n}}^2\!,\! \cdots \!,\!\sigma _{{G_M} \!\to\! {V_n}}^2} \right),
\end{equation}
\begin{equation}\tag{32}
{{\bf{Q}}_{{{\bf{n}}_{{V_n} \!\to\! V}}}} = {\mathop{\rm diag}\nolimits} \left( {\sigma _{{V_n} \!\to\! {V_{{L_{n,1}}}}}^2, \cdots ,\sigma _{{V_n} \!\to\! {V_{{L_{n,N - 1}}}}}^2} \right),
\end{equation}
and ${{\bf{I}}_{M - 1}}$ denotes the $\left( {M - 1} \right) \times \left( {M - 1} \right)$ identity matrix.

Then, the CRLB for UAV location estimation can be calculated using the following equation:
\begin{equation}\tag{33}
\begin{split}
{\mathop{\rm CRLB}\nolimits} \left( {\bf{v}} \right) &=  - E{\left[ {\frac{{{\partial ^2}\ln f\left( {{{\bf{o}}_V};{\bf{v}}} \right)}}{{\partial {\bf{v}}\partial {{\bf{v}}^T}}}} \right]^{ - 1}}\\
&=  - E{\left[\! {\frac{{{\partial ^2}\ln f\left( {{{\bf{d}}_{G \to V}};{\bf{v}}} \!\right)}}{{\partial {\bf{v}}\partial {{\bf{v}}^T}}} \!+\! \frac{{{\partial ^2}\ln f\left(\! {{{\bf{r}}_{G \to G}};{\bf{v}}} \!\right)}}{{\partial {\bf{v}}\partial {{\bf{v}}^T}}}} \!\right]^{ \!-\! 1}}\\
&= \left[ {\underbrace {{{\left( {\frac{{\partial {\bf{d}}_{G \to V}^ * }}{{\partial {\bf{v}}}}} \right)}^T}{\bf{Q}}_{{{\bf{n}}_{G \to V}}}^{ - 1}\left( {\frac{{\partial {\bf{d}}_{G \to V}^ * }}{{\partial {\bf{v}}}}} \right)}_{{\bf{J}}_{G \to V}^{\rm TDoA}}} \right.\\
&+ {\left. {\underbrace {{{\left( {\frac{{\partial {\bf{r}}_{V \to V}^ * }}{{\partial {\bf{v}}}}} \right)}^T}{\bf{Q}}_{{{\bf{n}}_{V \to V}}}^{ - 1}\left( {\frac{{\partial {\bf{r}}_{V \to V}^ * }}{{\partial {\bf{v}}}}} \right)}_{{\bf{J}}_{V \to V}^{\rm DR - TWR}}} \right]^{ - 1}},
\end{split}
\end{equation}
where ${\bf{J}}_{G \to V}^{{\rm{TDoA}}}$ and ${\bf{J}}_{V \to V}^{{\rm{DR - TWR}}}$ are Fisher information matrices (FIM) corresponding to G2V TDoA measurements and V2V DR-TWR measurements, respectively. It can be clearly seen from the above equation that the CRLB is the inverse of the sum of FIM ${\bf{J}}_{G \to V}^{{\rm{TDoA}}}$ and ${\bf{J}}_{V \to V}^{{\rm{DR - TWR}}}$, which reflects the contribution of the two types of measurements to position accuracy. $\frac{{\partial {\bf{d}}_{G \to V}^ * }}{{\partial {\bf{v}}}}$ in equation (33) is the partial derivative of TDoA measurements with respect to the UAVs' locations, which can be expressed as
\begin{equation}\tag{34}
\frac{{\partial {\bf{d}}_{G \to V}^ * }}{{\partial {\bf{v}}}} = {\mathop{\rm blkdiag}\nolimits} \left( {{{\bf{D}}_1},{{\bf{D}}_2}, \cdots ,{{\bf{D}}_N}} \right),
\end{equation}
where ${{\bf{D}}_n}$ is a $\left( {M - 1} \right) \times 2$ matrix, whose expression can be written as ($1 \le i \le M - 1$)
\begin{equation}\tag{35}
{{\bf{D}}_n}\left( {i,:} \right) = \frac{{\partial d_{{G_{i + 1}},{G_1} \to {V_n}}^ * }}{{\partial {{\bf{v}}_n}}} = {\left[ {{\bf{k}}_{{{\bf{g}}_{i + 1}}}^{{\bf{v}}_n^ * } - {\bf{k}}_{{{\bf{g}}_1}}^{{\bf{v}}_n^ * }} \right]^T},
\end{equation}
\begin{equation}\tag{36}
{\bf{k}}_{{{\bf{g}}_m}}^{{\bf{v}}_n^ * } = \frac{{\left( {{\bf{v}}_n^ *  - {{\bf{g}}_m}} \right)}}{{r_{{G_m} \to {V_n}}^ * }} = \frac{{\left( {{\bf{v}}_n^ *  - {{\bf{g}}_m}} \right)}}{{\left\| {{\bf{v}}_{n,3D}^ *  - {{\bf{g}}_{m,3D}}} \right\|}}.
\end{equation}
Similarly, the partial derivative $\frac{{\partial {\bf{r}}_{V \to V}^ * }}{{\partial {\bf{v}}}}$ of DR-TWR measurements can be expressed as equations (37)-(39).
\begin{equation}\tag{37}
\frac{{\partial {\bf{r}}_{V \to V}^ * }}{{\partial {\bf{v}}}} = {\left[ {{\bf{R}}_1^T,{\bf{R}}_2^T, \cdots ,{\bf{R}}_N^T} \right]^T},
\end{equation}
\begin{equation}\tag{39}
{\bf{k}}_{{\bf{v}}_{{L_{n,i}}}^ * }^{{\bf{v}}_n^ * } = \frac{{\left( {{\bf{v}}_n^ *  - {\bf{v}}_{{L_{n,i}}}^ * } \right)}}{{r_{{V_n} \to {V_{{L_{n,i}}i}}}^ * }} = \frac{{\left( {{\bf{v}}_n^ *  - {\bf{v}}_{{L_{n,i}}}^ * } \right)}}{{\left\| {{\bf{v}}_{n,3D}^ *  \!-\! {\bf{v}}_{{L_{n,i}},3D}^ * } \right\|}}.
\end{equation}

We assume that the UAV self-localization performed at GRS ${G_1}$ could achieve the CRLB. Then, the covariance matrix of the UAV position uncertainty can be approximated as
\begin{equation}\tag{40}
{{\bf{Q}}_{\Delta {\bf{v}}}} \!=\! {\mathop{\rm cov}} \left(\! {\Delta {\bf{v}}} \!\right) \!=\! E\left[ {\left(\! {{\bf{\hat v}} \!-\! {{\bf{v}}^ * }} \!\right){{\left(\! {{\bf{\hat v}} \!-\! {{\bf{v}}^ * }} \!\right)}^T}} \right] \!\approx\! {\mathop{\rm CRLB}\nolimits} \left( {\bf{v}} \right),
\end{equation}
where ${{\bf{v}}^ * } = {\left[ {{{\left( {{\bf{v}}_1^ * } \right)}^T}, \cdots ,{{\left( {{\bf{v}}_N^ * } \right)}^T}} \right]^T}$, $\Delta {\bf{v}} = {\left[ {\Delta {\bf{v}}_1^T\!,\! \cdots \!,\!\Delta {\bf{v}}_N^T} \right]^T}$.

\subsection{RMSE of UE Position Estimation}
As described in Section II.C, the UE uses the ILS method to determine its own location. Since the position and clock uncertainty of UAVs are unknown, the UE can only use the following measurement equations for position estimation:
\begin{equation}\tag{41}
{{\bf{\bar d}}_{V \to U}}\left( {\bf{u}} \right) = {\left[ {{{\bar d}_{{V_2},{V_1} \to U}}\left( {\bf{u}} \right), \cdots ,{{\bar d}_{{V_N},{V_1} \to U}}\left( {\bf{u}} \right)} \right]^T},
\end{equation}
\begin{equation}\tag{42}
{\bar d_{{V_n},{V_1} \to U}}\left( {\bf{u}} \right) = \left\| {{{\bf{u}}_{3D}} - {{{\bf{\hat v}}}_{n,3D}}} \right\| - \left\| {{{\bf{u}}_{3D}} - {{{\bf{\hat v}}}_{1,3D}}} \right\|,
\end{equation}
where ${{\bf{u}}_{3D}} = {\left[ {{{\bf{u}}^T},{h_U}} \right]^T}$.

The ILS method estimates the UE's location in an iterative manner through Taylor-series linearization \cite{ILS_Method}. Let ${{\bf{\hat u}}_k} = {\left[ {\hat x_U^k,\hat y_U^k} \right]^T}$ (${{\bf{\hat u}}_{k,3D}} = {\left[ {{{{\bf{\hat u}}}_k},{h_U}} \right]^T}$) denote the location estimate obtained in the \textit{k}-th iteration, the first-order Taylor-series expansion of ${{\bf{\bar d}}_{V \to U}}\left( {\bf{u}} \right)$ at ${\bf{u}} = {{\bf{\hat u}}_k}$ can be expressed as
\begin{equation}\tag{43}
{{\bf{\bar d}}_{V \to U}}\left( {\bf{u}} \right) \simeq {{\bf{\bar d}}_{V \to U}}\left( {{{{\bf{\hat u}}}_k}} \right) + {\bf{H}}\left( {{{{\bf{\hat u}}}_k}} \right)\left( {{\bf{u}} - {{{\bf{\hat u}}}_k}} \right),
\end{equation}
where
\begin{equation}\tag{44}
\begin{split}
{\bf{H}}\left( {{{{\bf{\hat u}}}_k}} \right) &= {\left. {\frac{{\partial {{{\bf{\bar d}}}_{V \to U}}\left( {\bf{u}} \right)}}{{\partial {\bf{u}}}}} \right|_{{{{\bf{\hat u}}}_k}}}\\
&= {\left[ {\left( {{\bf{k}}_{{{{\bf{\hat v}}}_2}}^{{{{\bf{\hat u}}}_k}} - {\bf{k}}_{{{{\bf{\hat v}}}_1}}^{{{{\bf{\hat u}}}_k}}} \right), \cdots ,\left( {{\bf{k}}_{{{{\bf{\hat v}}}_N}}^{{{{\bf{\hat u}}}_k}} - {\bf{k}}_{{{{\bf{\hat v}}}_1}}^{{{{\bf{\hat u}}}_k}}} \right)} \right]^T}\\
&\approx {\left[ {\left( {{\bf{k}}_{{\bf{v}}_2^ * }^{{{{\bf{\hat u}}}_k}} - {\bf{k}}_{{\bf{v}}_1^ * }^{{{{\bf{\hat u}}}_k}}} \right), \cdots ,\left( {{\bf{k}}_{{\bf{v}}_N^ * }^{{{{\bf{\hat u}}}_k}} - {\bf{k}}_{{\bf{v}}_1^ * }^{{{{\bf{\hat u}}}_k}}} \right)} \right]^T},
\end{split}
\end{equation}
is the Jacobian matrix of measurement equations at ${{\bf{\hat u}}_k}$, and
\begin{equation}\tag{45}
{\bf{k}}_{{\bf{v}}_n^ * }^{{{{\bf{\hat u}}}_k}} = \frac{{\left( {{{{\bf{\hat u}}}_k} - {\bf{v}}_n^ * } \right)}}{{\left\| {{{{\bf{\hat u}}}_{k,3D}} - {\bf{v}}_{n,3D}^ * } \right\|}}.
\end{equation}
Noted that in equation (44), the estimated locations (${{\bf{\hat v}}_n}$) of UAVs are replaced by the corresponding true values (${\bf{v}}_n^ * $). The reason and rationality for this operation were explained in \cite{APU_TDoA_1}.

Then, the least-squares estimate of the UE's location obtained in the (\textit{k+1})-th iteration is given by
\begin{equation}\tag{46}
\begin{split}
{{{\bf{\hat u}}}_{k \!+\! 1}} &= {{{\bf{\hat u}}}_k} \!+\! {\left( {{\bf{H}}{{\left(\! {{{{\bf{\hat u}}}_k}} \!\right)}^T}{\bf{Q}}_{{{\bf{n}}_{V \!\to\! U}}}^{ \!-\! 1}{\bf{H}}\left(\! {{{{\bf{\hat u}}}_k}} \!\right)} \right)^{ \!-\! 1}}{\bf{H}}{\left( {{{{\bf{\hat u}}}_k}} \right)^T}{\bf{Q}}_{{{\bf{n}}_{V \!\to\! U}}}^{ \!-\! 1}\\
&\quad\;\cdot \left( {{{\bf{d}}_{V \to U}} - {{{\bf{\bar d}}}_{V \to U}}\left( {{{{\bf{\hat u}}}_k}} \right)} \right),
\end{split}
\end{equation}
Replace ${{\bf{\hat u}}_k}$ in the above equation with the UE's true location ${{\bf{u}}^ * }$, the estimation error of the ILS method after convergence can be written as
\begin{equation}\tag{47}
\Delta {\bf{u}} \!=\! {\bf{\hat u}} - {{\bf{u}}^ * } \!=\! {\bf{S}}\left( {{{\bf{u}}^ * }} \right)\left( {{{\bf{d}}_{V \to U}} \!-\! {{{\bf{\bar d}}}_{V \to U}}\left( {{{\bf{u}}^ * }} \right)} \right),
\end{equation}
where
\begin{equation}\tag{48}
{\bf{S}}\left( {{{\bf{u}}^ * }} \right) = {\bf{P}}\left( {{{\bf{u}}^ * }} \right){\bf{H}}{\left( {{{\bf{u}}^ * }} \right)^T}{\bf{Q}}_{{{\bf{n}}_{V \to U}}}^{ - 1},
\end{equation}
\begin{equation}\tag{49}
{\bf{P}}\left( {{{\bf{u}}^ * }} \right) = {\left( {{\bf{H}}{{\left( {{{\bf{u}}^ * }} \right)}^T}{\bf{Q}}_{{{\bf{n}}_{V \to U}}}^{ - 1}{\bf{H}}\left( {{{\bf{u}}^ * }} \right)} \right)^{ - 1}};
\end{equation}
${{\bf{Q}}_{{{\bf{n}}_{V \!\to\! U}}}}$ denotes the covariance matrix of the noise term ${{\bf{n}}_{V \!\to\! U}}$ in equation (21), and can be expressed as
\begin{equation}\tag{50}
{{\bf{Q}}_{{{\bf{n}}_{V \!\to\! U}}}} \!=\! \sigma _{{V_1} \!\to\! U}^2 \!\cdot\! {{\bf{I}}_{N - 1}} \!+\! {\mathop{\rm diag}\nolimits} \left( {\sigma _{{V_2} \!\to\! U}^2, \cdots ,\sigma _{{V_N} \to U}^2} \right).
\end{equation}

We then derive the expression of term $\left(\! {{{\bf{d}}_{V \!\to\! U}} \!-\! {{{\bf{\bar d}}}_{V \!\to\! U}}\left(\! {{{\bf{u}}^ * }} \!\right)} \!\right)$ in equation (47). The difference between the measured TDoA ${d_{{V_n},{V_1} \to U}}$ and the TDoA ${\bar d_{{V_n},{V_1} \to U}}\left( {{{\bf{u}}^ * }} \right)$ predicted based on equation (42) can be written as
\begin{equation}\tag{51}
\begin{split}
{d_{{V_n},{V_1} \!\to\! U}} \!-\! {{\bar d}_{{V_n},{V_1} \!\to\! U}}\left( {{{\bf{u}}^ * }} \right) = \left( {d_{{V_n},{V_1} \!\to\! U}^ *  \!-\! \Delta t_{{V_n},{V_1} \!\to\! U}^{Pos}} \right.\\
- \Delta t_{{V_n},{V_1} \to U}^{Noi} + \left. {{n_{{V_n},{V_1} \to U}}} \right)\qquad\quad\quad\,\\
- \left( {\left\| {{\bf{u}}_{3D}^ *  \!-\! {{{\bf{\hat v}}}_{n,3D}}} \right\| - \left\| {{\bf{u}}_{3D}^ *  \!-\! {{{\bf{\hat v}}}_{1,3D}}} \right\|} \right),
\end{split}
\end{equation}
where the expression of $\Delta t_{{V_n},{V_1} \to U}^{Pos}$ is
\begin{equation}\tag{52}
\begin{split}
\Delta t_{{V_n},{V_1} \to U}^{Pos} &= \left( {\left\| {{{{\bf{\hat v}}}_{n,3D}} \!-\! {{\bf{g}}_{1,3D}}} \right\| - \left\| {{\bf{v}}_{n,3D}^ *  \!-\! {{\bf{g}}_{1,3D}}} \right\|} \right)\\
&- \left( {\left\| {{{{\bf{\hat v}}}_{1,3D}} \!-\! {{\bf{g}}_{1,3D}}} \right\| - \left\| {{\bf{v}}_{1,3D}^ *  \!-\! {{\bf{g}}_{1,3D}}} \right\|} \right).
\end{split}
\end{equation}
Expanding the term $\left\| {{{{\bf{\hat v}}}_{n,3D}} - {{\bf{g}}_{1,3D}}} \right\|$ at ${\bf{v}}_{n,3D}^ * $ based on the relationship ${{\bf{\hat v}}_{n,3D}} = {\bf{v}}_{n,3D}^ *  + {\left[ {\Delta {\bf{v}}_n^T,0} \right]^T}$, we have
\begin{equation}\tag{53}
\left\| {{{{\bf{\hat v}}}_{n,3D}} \!-\! {{\bf{g}}_{1,3D}}} \right\| \simeq \left\| {{\bf{v}}_{n,3D}^ *  \!-\! {{\bf{g}}_{1,3D}}} \right\| + {\left( {{\bf{k}}_{{{\bf{g}}_1}}^{{\bf{v}}_n^ * }} \right)^T}\Delta {{\bf{v}}_n}.
\end{equation}
Thus, $\Delta t_{{V_n},{V_1} \to U}^{Pos}$ in equation (51) can be approximated as
\begin{equation}\tag{54}
\Delta t_{{V_n},{V_1} \to U}^{Pos} \approx {\left( {k_{{{\bf{g}}_1}}^{{\bf{v}}_n^ * }} \right)^T}\Delta {{\bf{v}}_n} - {\left( {{\bf{k}}_{{{\bf{g}}_1}}^{{\bf{v}}_1^ * }} \right)^T}\Delta {{\bf{v}}_1}.
\end{equation}
Similarly, the term $\left\| {{\bf{u}}_{3D}^ *  \!-\! {{{\bf{\hat v}}}_{n,3D}}} \right\|$ in equation (51) can be expanded as
\begin{equation}\tag{54}
\left\| {{\bf{u}}_{3D}^ *  \!-\! {{{\bf{\hat v}}}_{n,3D}}} \right\| \simeq \left\| {{\bf{u}}_{3D}^ *  \!-\! {\bf{v}}_{n,3D}^ * } \right\| - {\left( {{\bf{k}}_{{\bf{v}}_n^ * }^{{{\bf{u}}^ * }}} \right)^T}\Delta {{\bf{v}}_n},
\end{equation}
and ${\bar d_{{V_n},{V_1} \to U}}\left( {{{\bf{u}}^ * }} \right)$ can be approximated as
\begin{equation}\tag{55}
\begin{split}
{{\bar d}_{{V_n},{V_1} \!\to\! U}}\left(\! {{{\bf{u}}^ * }} \!\right) &\approx\! \left(\! {\left\| {{\bf{u}}_{3D}^ *  \!-\! {\bf{v}}_{n,3D}^ * } \right\| \!-\! {{\left(\! {{\bf{k}}_{{\bf{v}}_n^ * }^{{{\bf{u}}^ * }}} \!\right)}^T}\!\Delta {{\bf{v}}_n}} \!\right)\\
&-\! \left(\! {\left\| {{\bf{u}}_{3D}^ *  \!-\! {\bf{v}}_{1,3D}^ * } \right\| \!-\! {{\left(\! {{\bf{k}}_{{\bf{v}}_1^ * }^{{{\bf{u}}^ * }}} \!\right)}^T}\!\Delta {{\bf{v}}_1}} \!\right)\\
&=\! d_{{V_n},{V_1} \!\to\! U}^ *  \!-\! {\left(\! {{\bf{k}}_{{\bf{v}}_n^ * }^{{{\bf{u}}^ * }}} \!\right)^T}\!\Delta {{\bf{v}}_n} \!+\! {\left(\! {{\bf{k}}_{{\bf{v}}_1^ * }^{{{\bf{u}}^ * }}} \!\right)^T}\!\Delta {{\bf{v}}_1}.
\end{split}
\end{equation}
Then, the expression of ${d_{{V_n},{V_1} \to U}} \!-\! {\bar d_{{V_n},{V_1} \to U}}\left( {{{\bf{u}}^ * }} \right)$ can be written as
\begin{equation}\tag{56}
\begin{split}
d{}_{{V_n},{V_1} \!\to\! U} \!-\! \bar d{}_{{V_n},{V_1} \!\to\! U}\left(\! {{{\bf{u}}^ * }} \!\right) \!=\! {\left(\! {{\bf{k}}_{{\bf{v}}_n^ * }^{{{\bf{u}}^ * }} \!-\! {\bf{k}}_{{{\bf{g}}_1}}^{{\bf{v}}_n^ * }} \!\right)^T}\!\Delta {{\bf{v}}_n}\qquad\quad\\
\qquad\qquad\!-\! {\left(\! {{\bf{k}}_{{\bf{v}}_1^ * }^{{{\bf{u}}^ * }} \!-\! {\bf{k}}_{{{\bf{g}}_1}}^{{\bf{v}}_1^ * }} \!\right)^T}\!\Delta {{\bf{v}}_1} \!-\! \Delta t_{{V_n},{V_1} \!\to\! U}^{Noi} \!+\! {n_{{V_n},{V_1} \!\to\! U}}.
\end{split}
\end{equation}
Moreover, the term $\left( {{{\bf{d}}_{V \to U}} \!-\! {{{\bf{\bar d}}}_{V \to U}}\left( {{{\bf{u}}^ * }} \right)} \right)$ in equation (47) can be expressed as
\begin{equation}\tag{57}
{{\bf{d}}_{V \!\to\! U}} \!-\! {{\bf{\bar d}}_{V \!\to\! U}}\left(\! {{{\bf{u}}^ * }} \!\right) = {{\bf{K}}_{V \!\to\! U}}\Delta {\bf{v}} \!-\! \Delta {\bf{t}}_{V \!\to\! U}^{Noi} \!+\! {{\bf{n}}_{V \!\to\! U}},
\end{equation}
where ${{\bf{K}}_{V \to U}}$ is a $\left( {N - 1} \right) \times 2N$ matrix and its expression can be written as ($1 \le i \le N - 1$)
\begin{equation}\tag{58}
\begin{split}
{{\bf{K}}_{V \to U}}\left( {i,:} \right) = \left[ { - {{\left( {{\bf{k}}_{{\bf{v}}_1^ * }^{{{\bf{u}}^ * }} - {\bf{k}}_{{{\bf{g}}_1}}^{{\bf{v}}_1^ * }} \right)}^T},{\bf{0}}_{2\left( {i - 1} \right) \times 1}^T,} \right.\qquad\quad\;\;\;\\
\left. {{{\left( {{\bf{k}}_{{\bf{v}}_{i + 1}^ * }^{{{\bf{u}}^ * }} - {\bf{k}}_{{{\bf{g}}_1}}^{{\bf{v}}_{i + 1}^ * }} \right)}^T},{\bf{0}}_{2\left( {N - i - 1} \right) \times 1}^T} \right].
\end{split}
\end{equation}

With equations (47) and (57), the covariance matrix of UE position error is derived as
\begin{equation}\tag{59}
\begin{split}
{{\bf{Q}}_{\Delta {\bf{u}}}} &=\! {\mathop{\rm cov}} \left( {\Delta {\bf{u}}} \right) \!=\! E\!\left[ {\left(\! {{\bf{\hat u}} \!-\! {{\bf{u}}^ * }} \!\right){{\left(\! {{\bf{\hat u}} \!-\! {{\bf{u}}^ * }} \!\right)}^T}} \right]\\
&=\! E\!\left[ {{\bf{S}}\left(\! {{{\bf{u}}^ * }} \!\right)\left( {{{\bf{K}}_{V \!\to\! U}}\Delta {\bf{v}} \!-\! \Delta {\bf{t}}_{V \!\to\! U}^{Noi} \!+\! {{\bf{n}}_{V \!\to\! U}}} \!\right)} \!\right.\\
&\qquad\left. {{{\left( {{{\bf{K}}_{V \!\to\! U}}\Delta {\bf{v}} \!-\! \Delta {\bf{t}}_{V \!\to\! U}^{Noi} \!+\! {{\bf{n}}_{V \!\to\! U}}} \!\right)}^T}{\bf{S}}{{\left(\! {{{\bf{u}}^ * }} \!\right)}^T}} \right]\\
= &{\bf{P}}\!\left(\! {{{\bf{u}}^ * }} \!\right) +\! {\bf{S}}\left(\! {{{\bf{u}}^ * }} \!\right)\left(\! {{{\bf{K}}_{V \!\to\! U}}{{\bf{Q}}_{\Delta {\bf{v}}}}{\bf{K}}_{V \!\to\! U}^T \!+\! {{\bf{Q}}_{\Delta {\bf{t}}_{V \!\to\! U}^{Noi}}}} \!\right){\bf{S}}{\left(\! {{{\bf{u}}^ * }} \!\right)^T},
\end{split}
\end{equation}
where
\begin{equation}\tag{60}
{{\bf{Q}}_{\Delta {\bf{t}}_{V \!\to\! U}^{Noi}}} = \sigma _{{G_1} \!\to\! {V_1}}^2 \!\cdot\! {{\bf{I}}_{N \!-\! 1}} + {\mathop{\rm diag}\nolimits} \left( {\sigma _{{G_1} \!\to\! {V_2}}^2\!,\! \cdots \!,\!\sigma _{{G_1} \!\to\! {V_N}}^2} \right).
\end{equation}

Finally, the RMSE of UE position error can be calculated as follows:
\begin{equation}\tag{61}
{\mathop{\rm RMSE}\nolimits} \left( {\bf{u}} \right) = {\mathop{\rm trace}\nolimits} {\left( {{{\bf{Q}}_{\Delta {\bf{u}}}}} \right)^{\frac{1}{2}}}.
\end{equation}

As can be seen from equation (59), the covariance matrix ${{\bf{Q}}_{\Delta {\bf{u}}}}$ is the sum of two terms. The first of them (${\bf{P}}\left( {{{\bf{u}}^ * }} \right)$) represents the UE position error under ideal conditions where UAVs' locations are perfectly known and their clocks are accurately synchronized. The second term (${\bf{S}}\left(\! {{{\bf{u}}^ * }} \!\right)\left(\! {{{\bf{K}}_{V \!\to\! U}}{{\bf{Q}}_{\Delta {\bf{v}}}}{\bf{K}}_{V \!\to\! U}^T \!+\! {{\bf{Q}}_{\Delta {\bf{t}}_{V \!\to\! U}^{Noi}}}} \!\right){\bf{S}}{\left(\! {{{\bf{u}}^ * }} \!\right)^T}$) reflects the impacts of UAV position uncertainty and synchronization errors caused by noise and jamming on position accuracy.

\section{Numerical Results}
\begin{table}[!t]
\renewcommand{\arraystretch}{1.5}
\newcommand{\tabincell}[2]{\begin{tabular}{@{}#1@{}}#2\end{tabular}}
\caption{Simulation Parameters}
\label{table_comp}
\centering
\begin{tabular}{|l|l|}
\hline
Parameter&Value \\
\hline
Main frequency (${f_c}$)& 2.4 GHz \\
\hline
Signal bandwidth ($B$)& 10 MHz \\
\hline
Noise power (${P_{{n_0}}}$)& -95 dBm \\
\hline
Reference path loss at 1m (${\beta _0}$)& $1.01 \times {10^4}$ \\
\hline
PLE of G2A channels under LoS conditions ($\alpha _{G2A}^L$)& 2 \\
\hline
PLE of G2A channels under NLoS conditions ($\alpha _{G2A}^N$)& 3.2 \\
\hline
PLE of G2G channels under LoS conditions ($\alpha _{G2G}^L$)& 2.2 \\
\hline
Height of the jammer's antenna (${h_J}$)& 5 m \\
\hline
Transmit power of the jammer ($P_J^t$)& 20 dBm\\
\hline
Radius of the jamming area (${R_J}$)& 2.5 km \\
\hline
Number of GRSs ($M$)& 6 \\
\hline
Height of GRSs' antenna (${h_G}$)& 25 m \\
\hline
Transmit power of GRSs ($P_G^t$)& 35 dBm \\
\hline
Number of UAV ($N$)& 6 \\
\hline
UAV altitude (${h_V}$)& 100 m \\
\hline
Transmit power of UAVs ($P_V^t$)& 30 dBm \\
\hline
Height of the UE's antenna (${h_U}$)& 1.5 m \\
\hline
Side length of the target area (${L_T}$)& 500 m \\
\hline
\end{tabular}
\end{table}
In this section, a series of simulation experiments are conducted to evaluate the performance of the proposed system under jamming attacks, and the corresponding numerical results are presented to verify its feasibility and validity. First, we test and compare the position accuracy of the proposed UAV-assisted system and the conventional terrestrial positioning system using only GRSs in a typical jamming scenario. Then, the key factors affecting the anti-jamming performance of our system and their influence on position accuracy are analyzed in detail through several experiments. Table I summarizes the key simulation parameters used in this section.

Fig. 4 shows the jamming scenario for performance evaluation in this section, which consists of 6 GRSs, 6 UAVs and a jammer. The location of the jammer is set as the origin of coordinates, that is, ${\bf{w}} = {\left[ {0,0} \right]^T}$. GRSs are located on the boundary of the jamming area, and the difference between the azimuth angles of two adjacent GRSs is 20 degree. The horizontal coordinates of the 6 UAVs are set to ${\bf{v}}_1^ *  \!=\! {\left[ {1350, - 400} \right]^T}$, ${\bf{v}}_2^ *  \!=\! {\left[ {950,0} \right]^T}$, ${\bf{v}}_3^ *  \!=\! {\left[ {550, - 400} \right]^T}$, ${\bf{v}}_4^ *  \!=\! {\left[ {550,400} \right]^T}$, ${\bf{v}}_5^ *  \!=\! {\left[ {1750,0} \right]^T}$ and ${\bf{v}}_6^ *  \!=\! {\left[ {1350,400} \right]^T}$. The J2V links between UAVs and the jammer are dominated by LoS components. The target area where the UE is located is a square with center at ${{\bf{o}}_T} = {\left[ {950,0} \right]^T}$ and side length of 500m. During the simulation, the target area will be discretized into a series of sample points with an interval of 10m, and the position accuracy at each sample point would be calculated and recorded for performance evaluation.

\subsection{Feasibility and Validity of the Proposed System}
\begin{figure}[!t]
\centering
\includegraphics[height=1.90in,width=2.55in]{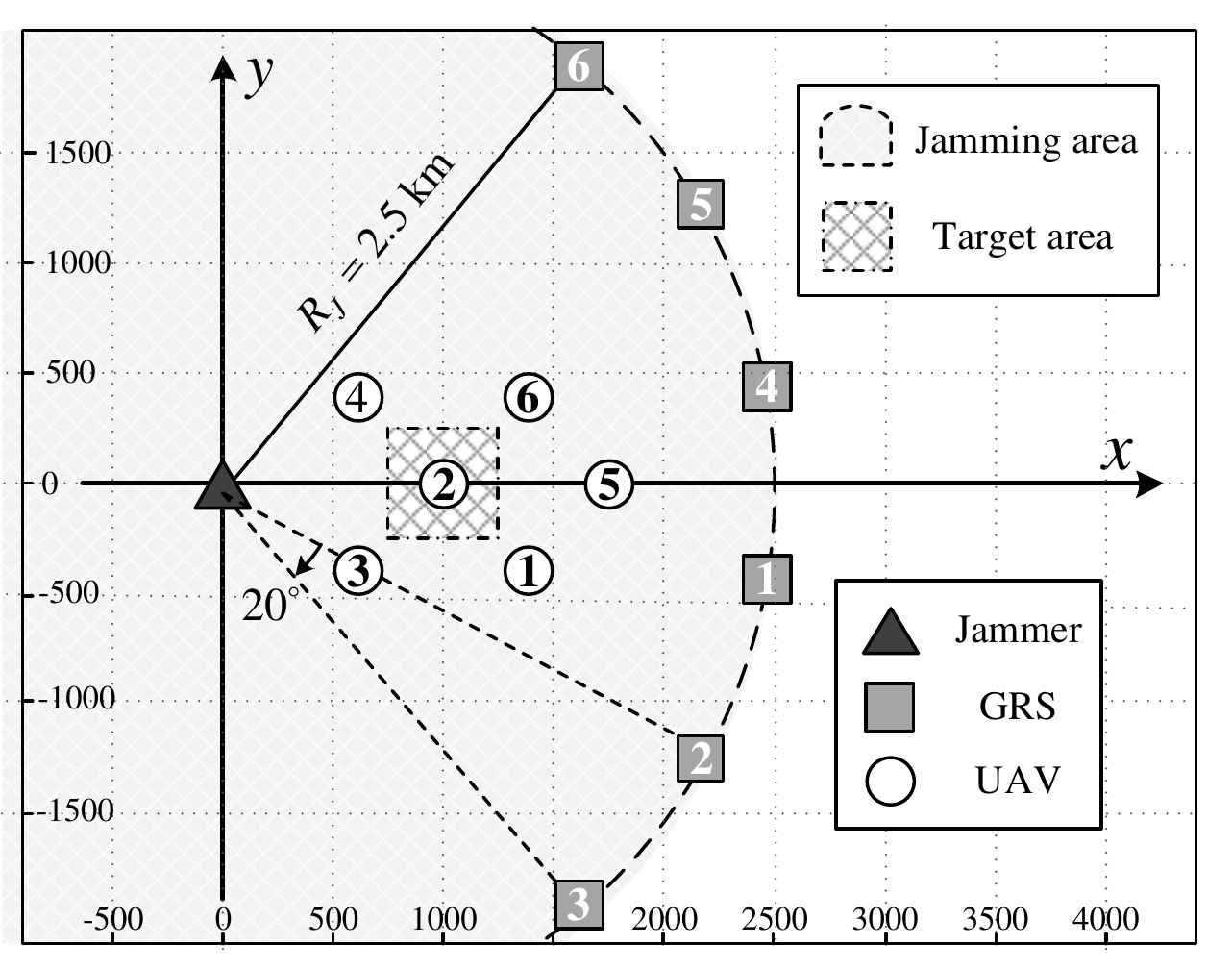}
\caption{Typical jamming scenario for numerical evaluation.}
\label{fig_4}
\end{figure}
\begin{figure}[!t]
\centering
\includegraphics[height=2.75in,width=3.45in]{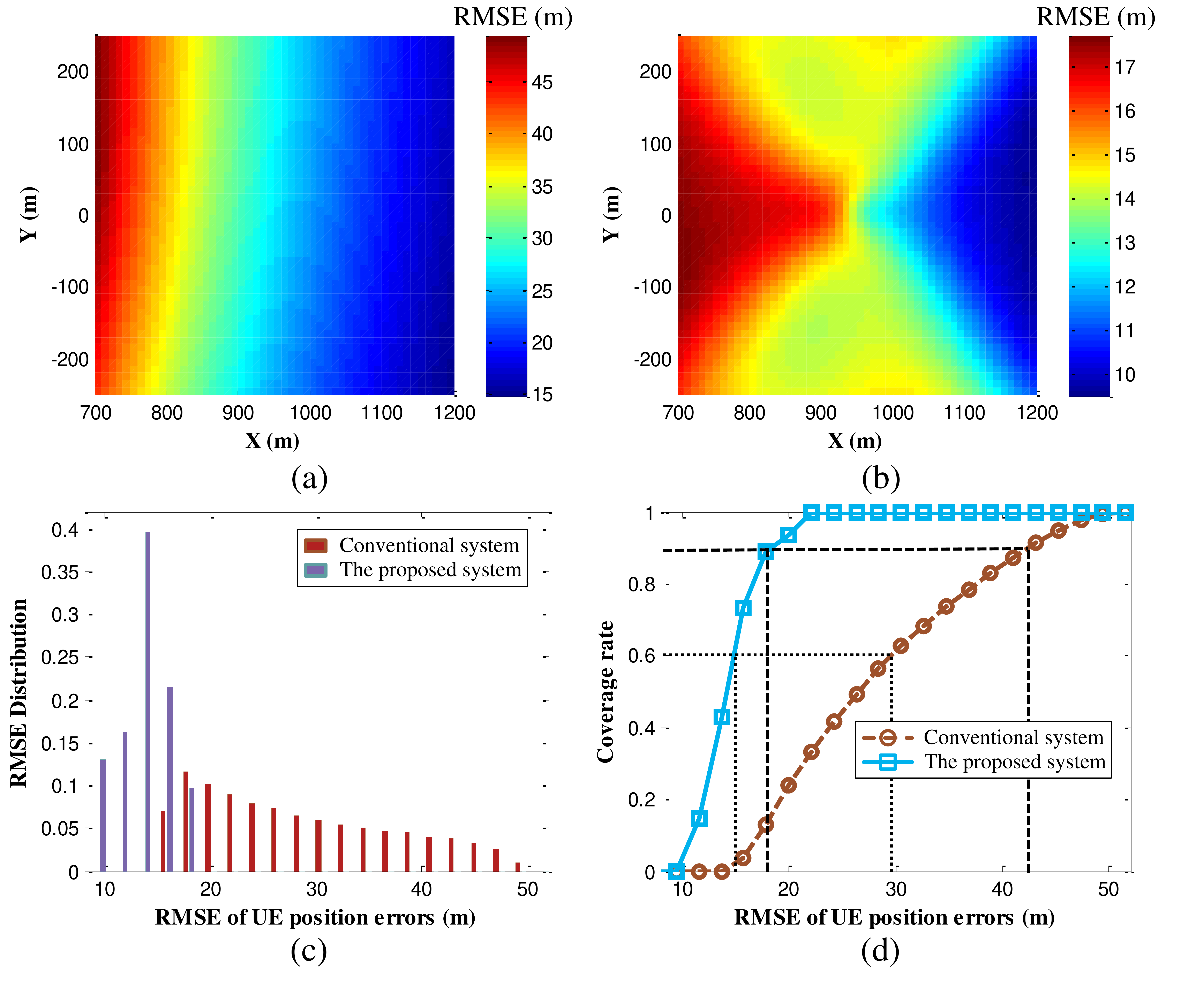}
\caption{Feasibility and performance of the proposed system: (a) UE position RMSE of the conventional (G2U LoS links exist) and (b) the proposed systems, (c) RMSE distributions and (d) service coverage of the two systems.}
\label{fig_5}
\end{figure}
With the expressions derived in Section III, we calculate the theoretical RMSE of UE position errors of the proposed system in the jamming scenario mentioned above, and compare it with that of the conventional terrestrial system using only GRSs. Noted that when talking about the conventional system, we assume that there are LoS paths between GRSs and the UE. The evaluation results of the two systems are shown in Fig. 5. Comparing the ``heat maps'' shown in Fig. 5(a) and (b), it can be found that the maximum RMSE of the proposed system in the target area is 17.7m, which is 64.2$\%$ lower than the 49.4m of the conventional system.

Moreover, in Fig. 5(c) and (d), we further analyze the RMSE distribution and service coverage rate of the two systems in the target area. As can be seen from Fig. 5(c), the conventional system's RMSE at most sample points is between 20m and 50m, much larger than the maximum RMSE of the proposed system. At only a small part of sample points, the RMSE of the conventional system is less than 20m, which is comparable to the proposed system's performance. The ``service coverage rate'' in Fig. 5(d) reflects the proportion of the areas with RMSE less than a certain value in the entire target area. For example, the conventional system's RMSE corresponding to the 90$\%$ coverage rate is 42.5m, which can be interpreted as: when using the conventional system, the positioning service with RMSE less than 42.5m could cover 90$\%$ of the target area. As can be seen from Fig. 5(d), the proposed system's 60$\%$ and 90$\%$ coverage RMSE is 14.9m and 18.5m, much smaller than the 29.7m and 42.5m of the conventional system. Therefore, it can be concluded from the above analysis that in jamming environments, the proposed UAV-assisted system outperforms the conventional terrestrial system in terms of maximum RMSE, RMSE distribution and service coverage, which demonstrate the feasibility and validity of our system.

The above evaluation results are quite reasonable. Due to the existence of the jamming area, the geometry of GRSs is unfavorable for positioning service, resulting in poor accuracy of the conventional terrestrial system. Thus, although their position and clock uncertainty will introduce additional errors in UE location estimation, UAVs with satisfactory geometry could still achieve better performance than the conventional system. Noted that this simulation experiment is based on the assumption that the propagation condition of G2G links is LoS, which may not be true in practice due to the long distance between GRSs and the UE. When there is no LoS path in G2G links, the conventional system cannot provide positioning services, while the proposed system can still work efficiently, which also reflects the superiority of our system.

\subsection{The Importance of V2V Links in the Proposed System}
\begin{figure}[!t]
\centering
\includegraphics[height=2.75in,width=3.45in]{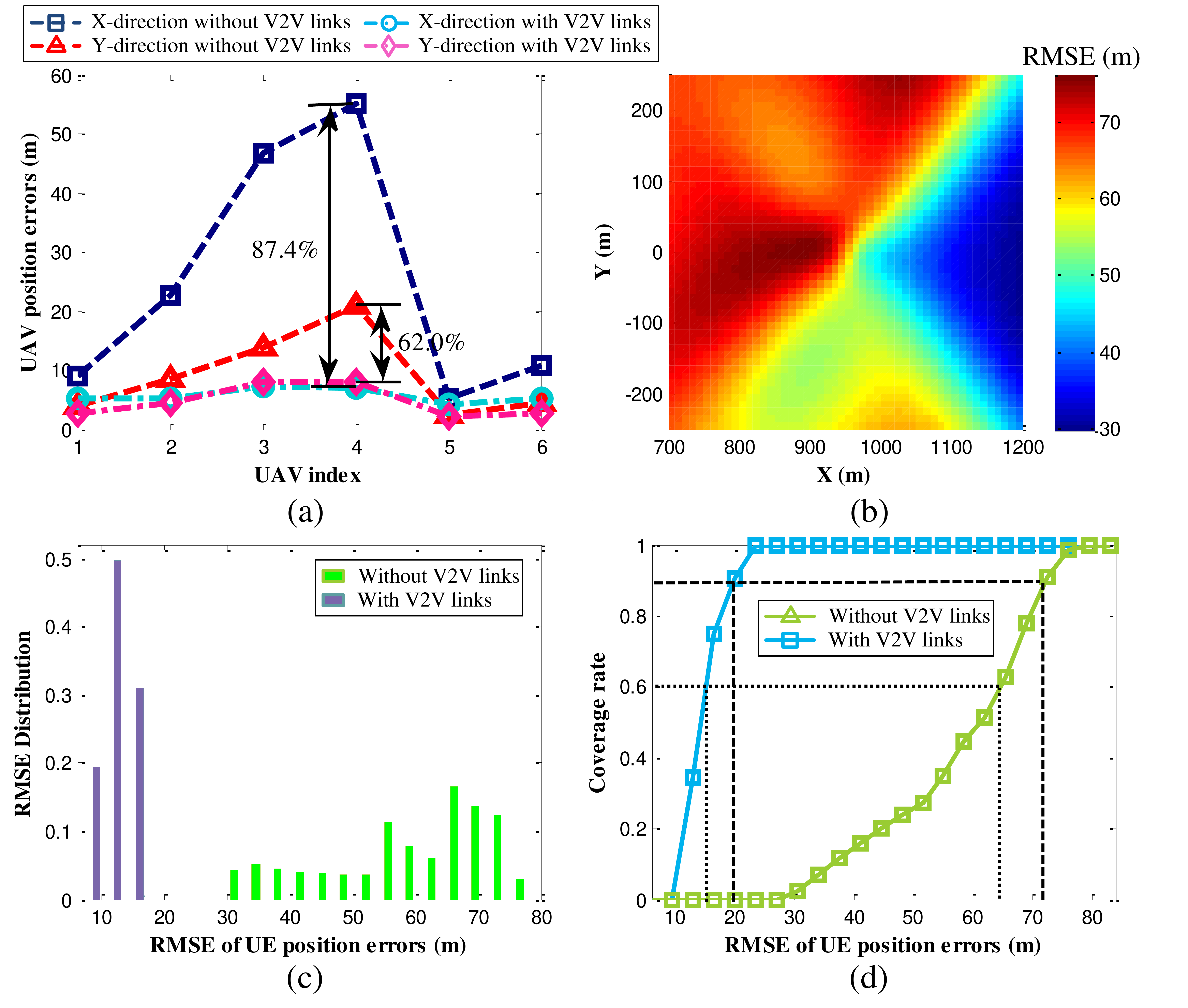}
\caption{Importance of V2V links: (a) UAV position errors under two conditions (with or without V2V links), (b) UE position RMSE without V2V links, (c) RMSE distributions and (d) service coverage under two conditions.}
\label{fig_6}
\end{figure}
As described in Section II.A, UAVs in the proposed system utilize the TDoA measurements from G2V links and the DR-TWR measurements from V2V links to determine their own locations. Obviously, the G2V link is indispensable for UAV self-localization as it provides valuable information about the UAVs' absolute locations. In contrast, it is not easy to recognize the role of the V2V measurement link in the proposed system. Thus, in this subsection, we carry out a simulation experiment to verify the importance of V2V links.

We first generate a simplified version of the proposed system by removing all V2V links, and then calculate its theoretical performance in the target area. The CRLB of the UAV self-localization without V2V links can be obtained by excluding the term ${\bf{J}}_{V \to V}^{{\rm{DR - TWR}}}$ in equation (33). Assign the newly calculated CRLB to the matrix ${{\bf{Q}}_{\Delta {\bf{v}}}}$ in equation (59), then the expressions derived in Section III.B can also be used to calculate the RMSE of UE position errors in the simplified system. The performance evaluation results of the simplified system are shown in Fig. 6.

It can be seen from Fig. 6(a) that the existence of V2V links reduces the maximum UAV position error in the x- and y-directions by 87.4$\%$ and 62.0$\%$, respectively. Intuitively, the reason for the huge improvement in the UAV position accuracy is that V2V links greatly reduce the impacts of GRSs' geometry on UAV self-localization. In the absence of V2V links, most of the GRSs are located on the right side of UAVs, resulting in poor geometry of anchor nodes, especially in the x-direction. With the V2V links, each UAV could be regarded as an anchor node for other UAVs, which improves the geometry and leads to satisfactory performance of UAV self-localization. As shown in Fig. 6(b), the maximum RMSE of the simplified system in the target area is 76.1m, about four times larger than the 17.7m of the proposed system (Fig. 5(b)) and even worse than that of the conventional terrestrial system (Fig. 5(a)). Moreover, it can be seen from Fig. 6(c) that the RMSE of the proposed system at all sample points is much smaller than the minimum RMSE of the simplified system without V2V links. Since the V2V links greatly reduce the UAV position uncertainty, it is natural that the UE position accuracy would be improved. In addition, according to Fig. 6(d), the simplified system's 60$\%$ and 90$\%$ coverage RMSE is 64.7m and 72.3m, which is significantly worse than that of the proposed system. All these phenomena demonstrate the importance of V2V links in the proposed UAV-assisted system.

\subsection{The Influence of J2V Links' Propagation Conditions}
\begin{figure}[!t]
\centering
\includegraphics[height=2.75in,width=3.45in]{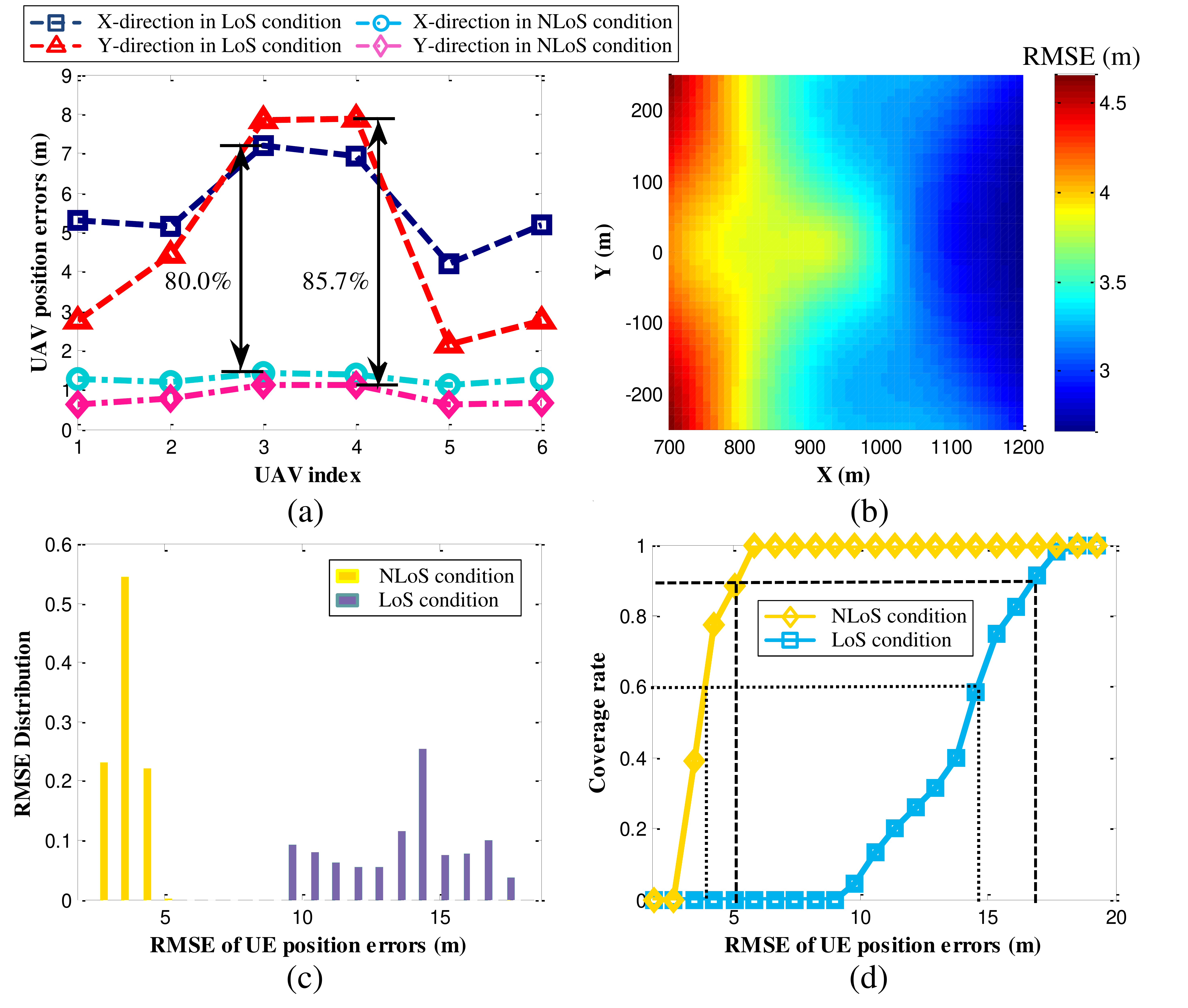}
\caption{Influence of NLoS propagation in J2V links: (a) UAV position errors under two conditions (J2V NLoS or LoS), (b) UE position RMSE when the propagation condition of J2V links is NLoS, (c) RMSE distribution and (d) service coverage under two conditions.}
\label{fig_7}
\end{figure}
Different from the UE, UAVs with high maneuverability are capable of changing the propagation conditions of J2V links through strategies like hiding behind buildings or mountains, so as to mitigate the impacts of jamming on positioning services. In order to investigate the influence of J2V links' propagation conditions on the proposed system's performance, we change the condition to NLoS and repeat the performance evaluation. The evaluation results are shown in Fig. 7.

It can be seen from Fig. 7(a) that in terms of UAV self-localization, changing the condition from LoS to NLoS reduces the maximum position error in x- and y-directions by 80.0$\%$ and 85.7$\%$, respectively. Since the change of the propagation condition greatly weakens the power of jamming signals received at UAVs, the reduction of UAV position error is not surprising. According to Fig. 7(b) and (c), the RMSE of UE position errors in the entire target area under NLoS conditions is less than 4.7m, which is only about half of the minimum RSME obtained under LoS conditions. Moreover, the curves in Fig. 7(d) show that the 60$\%$ and 90$\%$ coverage RMSE of the proposed system under NLoS conditions is 3.9m and 5.1m, which meets the requirements of meter-level positioning services.

The evaluation results in this subsection demonstrate that the NLoS propagation in J2V links is beneficial for improving the position accuracy of both the UAV and UE. Therefore, in practical applications, as long as the positioning services are not affected, UAVs should flexibly adjust their locations to avoid LoS paths between them and the jammer.

\subsection{The Influence of GRSs' Signal-to-Jammer Ratio}
\begin{figure}[!t]
\centering
\includegraphics[height=1.70in,width=3.45in]{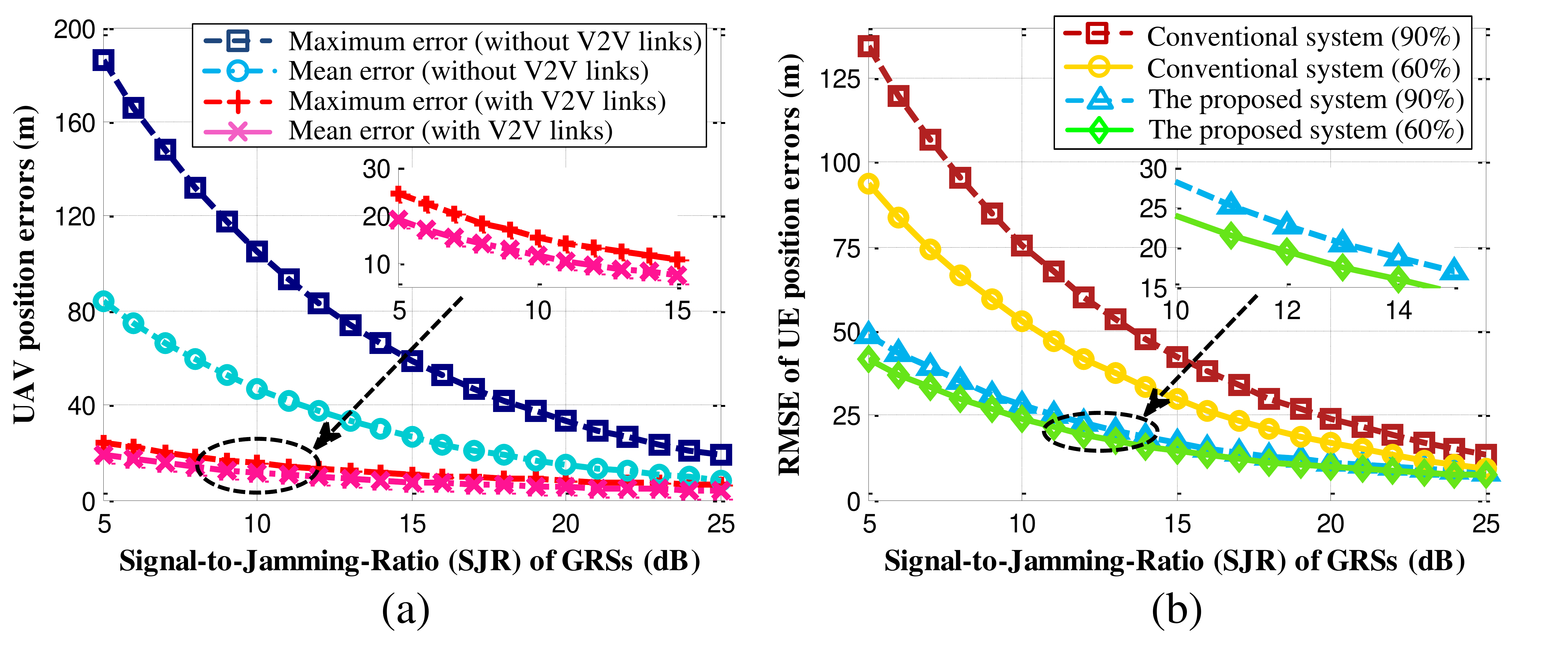}
\caption{Influence of GRSs' Signal-to-Jammer Ratio (SJR) on (a) UAV and (b) UE positioning performance.}
\label{fig_8}
\end{figure}
In this subsection, a simulation experiment is conducted to investigate the influence of GRSs' signal-to-jammer ratio (SJR) on the proposed system's performance. During the experiment, the transmit power of UAVs and the jammer is set according to Table I and remains unchanged, while GRSs' power increases from 25 to 45dBm (SJR: 5 to 25dB). We evaluate the performance of the proposed system, its simplified version without V2V links, and the conventional terrestrial system at each SJR. Both the CRLB of UAV self-localization and RMSE of UE position errors are calculated. The simulation results obtained are shown in Fig. 8.

It can be seen that the position accuracy of both the UAV and UE improves with the increase of GRSs' SJR. As shown in Fig. 8(a), in terms of UAV self-localization, increasing the SJR from 5 to 10dB reduces the maximum and mean UAV position errors in the proposed system by 37.4$\%$ and 39.6$\%$, respectively. Moreover, as the GRSs' SJR exceeds 20dB, the UAV position accuracy of the simplified system becomes very close to that of the proposed system. The explanation for this phenomenon is that the SINR of the V2V links does not change with GRSs' transmit power, so that the location information provided by V2V links is negligible when GRSs' SJR is extremely high.

According to Fig. 8(b), in terms of the positioning services provided for UE, as the SJR increases from 5 to 10dB, the 60$\%$ and 90$\%$ coverage RMSE of the proposed system is reduced by 42.4$\%$ and 42.5$\%$, respectively. Under the condition of ${\rm{SJR}} > 20{\rm{dB}}$, the RMSE of the conventional system is close to that of the proposed system. The reason for this phenomenon is quite similar to that explained in the previous paragraph, that is, due to their constant transmit power, UAVs no longer have advantages in anti-jamming after the GRSs' SJR exceeds a certain value. Of course, if there is no LoS path between GRSs and the UE, UAV would always be the best choice to provide positioning services for the target area regardless of GRSs' SJR. In addition, it can be found in Fig. 8(b) that the decreasing trend of the proposed system's RMSE slows down with the increase of SJR. This is because the UE position error caused by the noise and jamming in V2U links does not change with GRSs' SJR. If GRSs further increase their transmit power, the second term of equation (59) will approaches to 0, and the RMSE of the proposed system could be approximated by the first term ${\bf{P}}\left( {{{\bf{u}}^ * }} \right)$.

\subsection{The Influence of UAVs' Signal-to-Jammer Ratio}
\begin{figure}[!t]
\centering
\includegraphics[height=2.75in,width=3.45in]{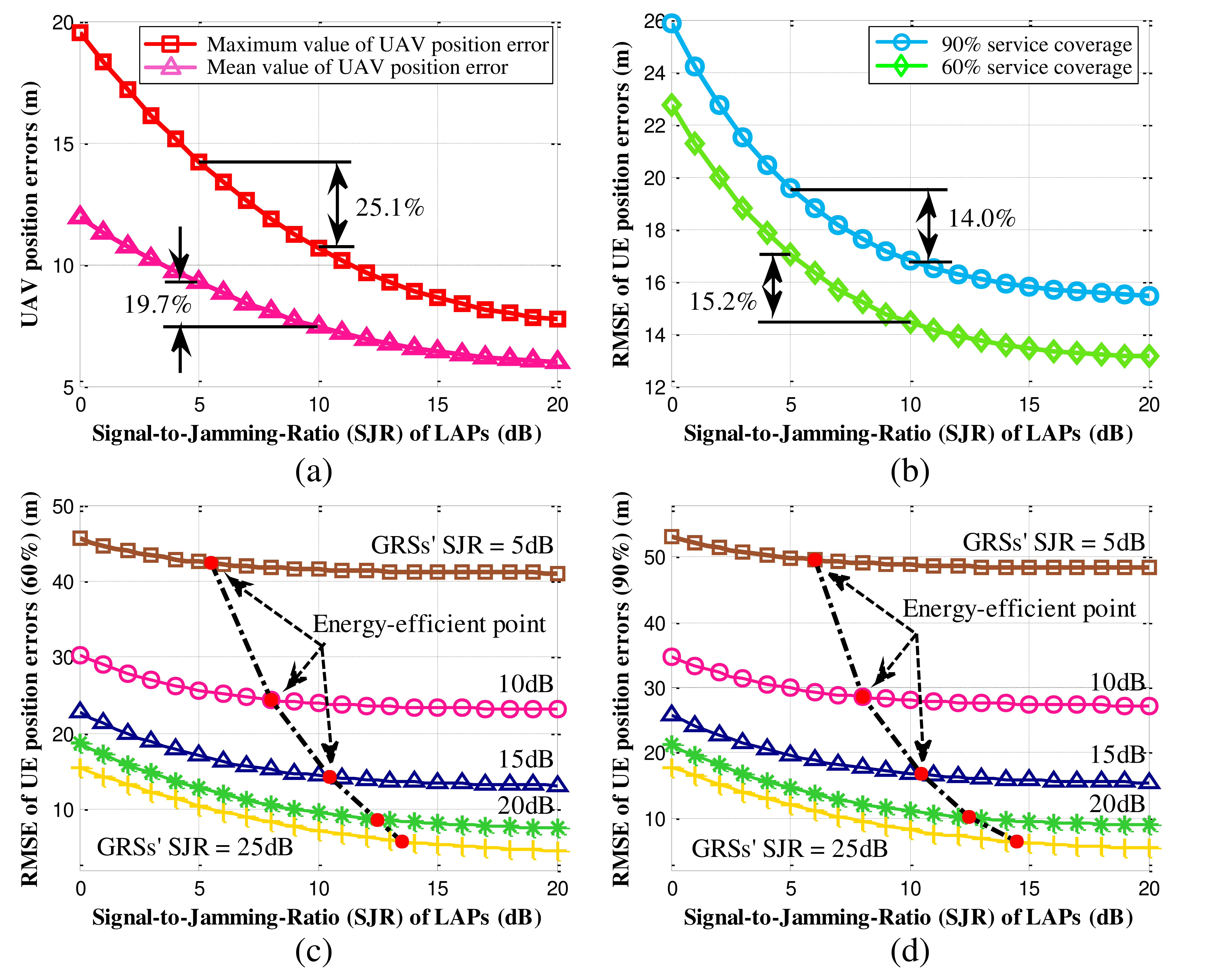}
\caption{Influence of UAVs' Signal-to-Jammer Ratio (SJR) on (a) UAV and (b) UE positioning performance, energy-efficient points for (c) 60$\%$ and (d) 90$\%$ service coverage.}
\label{fig_9}
\end{figure}
Unlike GRSs, UAVs with size and weight constraints commonly have very limited on-board energy, so their transmit power should be chosen carefully. In this subsection, we analyze the influence of UAVs' SJR on the proposed system's performance and try to find the transmit power with satisfactory position accuracy and energy efficiency. During the simulation, the transmit power of GRSs and the jammer remains constant as shown in Table I, while the UAVs' power varies from 20 to 40dBm (SJR: 0 to 20dB). The performance of the proposed system at each SJR is evaluated and shown in Fig. 9.

It can be seen from Fig. 9(a) and (b) that increasing the SJR of UAVs could significantly improve the performance of the UAV self-localization and UE positioning. According to Fig. 9(a), after the SJR increased from 5 to 10dB, the maximum and mean UAV position errors in the proposed system were reduced by 25.1$\%$ and 19.7$\%$, respectively. The 60$\%$ and 90$\%$ coverage RMSE of UE position errors was reduced by 15.2$\%$ and 14.0$\%$, as shown in Fig. 9(b). Moreover, it is noteworthy that the downward trends of the UAV and UE position errors slow down as the SJR increases. The explanation for this phenomenon is that the SINR of G2V links that provide absolute location information for UAVs does not change with UAVs' SJR. Thus, there is an upper limit to the improvement of the UAV and UE position accuracy brought about by the increase of SJR. It is unwise to continuously increase the UAVs' transmit power while the power of GRSs remains unchanged.

In order to avoid unnecessary wastage of on-board energy, we try to find the energy-efficient transmit power for UAVs in the proposed system. The curves in Fig. 9(c) and (d) show the RMSE of UE position errors obtained at different GRSs' SJR and UAVs' SJR. Our approach for power selection is as follows: gradually increase the UAVs' power from 20dBm. If the RMSE reduction caused by a further increase of 0.5dBm in power is less than 0.15m, then the current power is regarded as the energy-efficient transmit power for UAVs. The UAVs' SJR corresponding to the transmit power selected by this approach is indicated by the red dots in Fig. 9(c) and (d). For example, when the SJR of GRSs is 20dB, the UAVs' energy-efficient power for 60$\%$ service coverage is 30.5dBm (SJR: 10.5dB). The energy-efficient transmit power for 60$\%$ and 90$\%$ service coverage is slightly different (no more than 1dBm), and can be selected according to mission requirements in practice.

\section{Conclusion}
In this article, we presented a novel UAV-assisted anti-jamming positioning system that could provide services for users in jamming environments. In the proposed system, multiple low-altitude UAVs first utilize ground facilities to locate themselves, and then act as aerial anchor nodes to provide positioning services. We determined the structure and mathematical models of our system, and selected appropriate methods to support the UAV self-localization and positioning service. In order to evaluate the proposed system's theoretical performance, we further derive the CRLB for UAV self-localization and the RMSE of UE position estimate in the presence of jamming. In particular, the UAV position and clock uncertainty caused by jamming and noise are taken into account in the above derivation. Numerical results demonstrate the feasibility and validity of the proposed system, as well as its superior anti-jamming performance over the conventional terrestrial systems. We hope this article could lead to a new paradigm for the design of anti-jamming positioning systems.

\appendices
\section{Model of DR-TWR Technique}
\begin{figure}[!t]
\centering
\includegraphics[height=1.60in,width=2.90in]{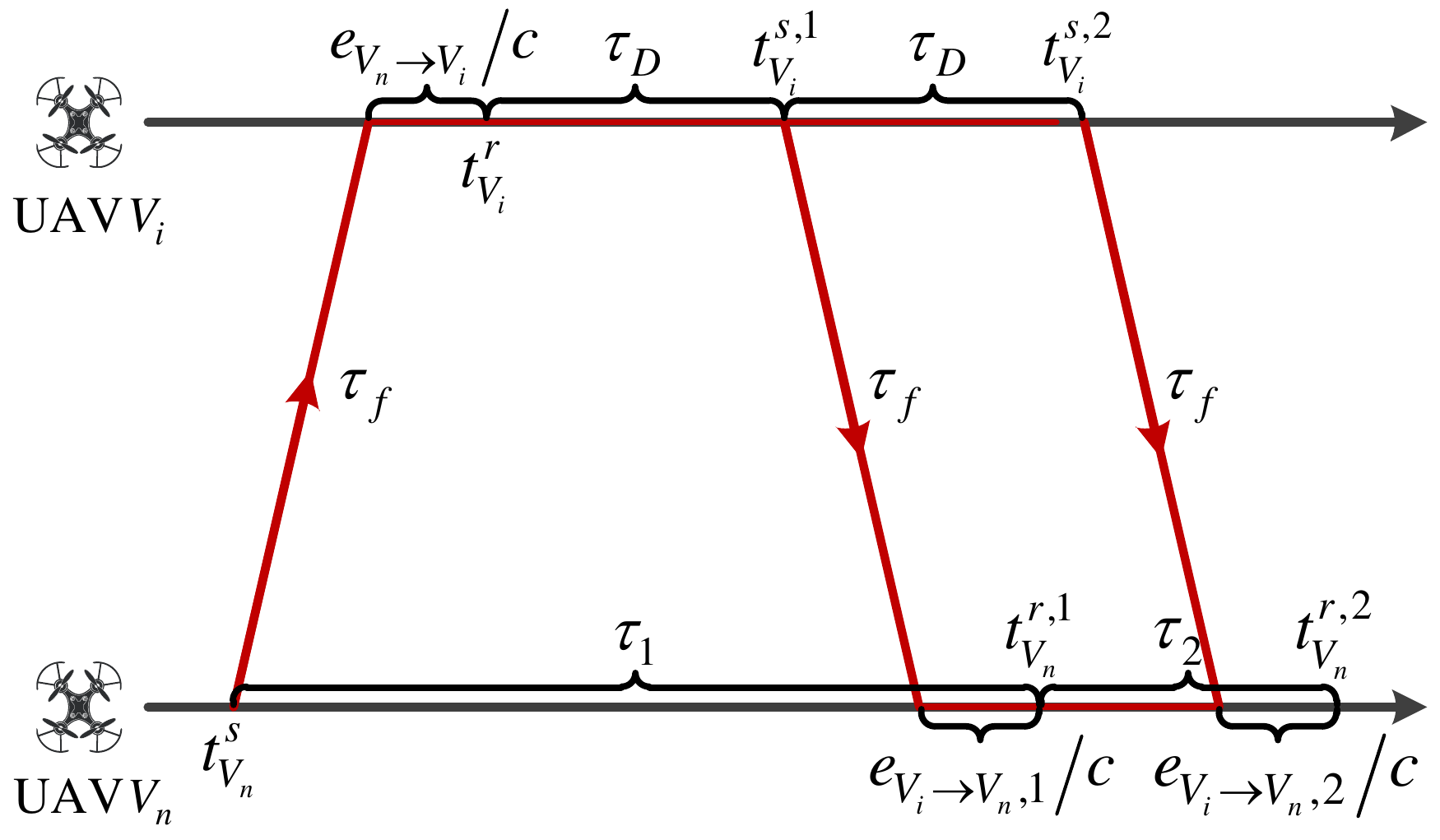}
\caption{Time diagram of DR-TWR.}
\label{fig_10}
\end{figure}
In this appendix, we introduce the DR-TWR protocol used in V2V links and derive its measurement model (equation (11)) \cite{DR_TWR}. As shown in Fig. 10, the DR-TWR corresponding to UAV pair $\left\langle {{V_n},{V_i}} \right\rangle $ ($i \ne n$) begins with a range request message send by UAV ${V_n}$. After detecting the request message, UAV ${V_i}$ first waits for ${\tau _D}$ seconds according to its local clock, and then sends back two response messages successively at an interval of ${\tau _D}$ seconds. UAV ${V_n}$ utilizes its local clock to measure the time interval ${\tau _1}$ between the transmission of request message ($t_{{V_n}}^s$) and the reception of the first response message ($t_{{V_n}}^{r,1}$), as well as ${\tau _2}$ between the receptions of two response messages ($t_{{V_n}}^{r,1}$ and $_{{V_n}}^{r,2}$). Denote the clock drifts of UAVs' local clocks relative to the reference clock as ${\delta _{{V_n}}}$ and ${\delta _{{V_1}}}$, then the measured time intervals (${\hat \tau _1}$ and ${\hat \tau _2}$) can be expressed as
\begin{equation}\tag{62}
\begin{split}
{{\hat \tau }_1} = t_{{V_n}}^{r,1} - t_{{V_n}}^s &= 2{\tau _f} \cdot \left( {1 + {\delta _{{V_n}}}} \right)\\
&+\! \left(\! {{\tau _D} \!+\! \frac{{{e_{{V_n} \!\to\! {V_i}}}}}{c}} \!\right) \!\cdot\! \frac{{\left(\! {1 \!+\! {\delta _{{V_i}}}} \!\right)}}{{\left(\! {1 \!+\! {\delta _{{V_n}}}} \!\right)}} \!+\! \frac{{{e_{{V_i} \!\to\! {V_n},1}}}}{c},
\end{split}
\end{equation}
\begin{equation}\tag{63}
{\hat \tau _2} = t_{{V_n}}^{r,2} \!-\! t_{{V_n}}^{r,1} = {\tau _D} \!\cdot\! \frac{{\left(\! {1 \!+\! {\delta _{{V_i}}}} \!\right)}}{{\left(\! {1 \!+\! {\delta _{{V_n}}}} \!\right)}} \!+\! \frac{{{e_{{V_i} \!\to\! {V_n},2}}}}{c} \!-\! \frac{{{e_{{V_i} \to {V_n},1}}}}{c},
\end{equation}
where ${e_{{V_n} \to {V_i}}} \sim {\cal N}\left( {0,\sigma _{{V_n} \to {V_i}}^2} \right)$, ${e_{{V_i} \to {V_n},1}} \sim {\cal N}\left( {0,\sigma _{{V_i} \to {V_n}}^2} \right)$ and ${e_{{V_i} \to {V_n},2}} \sim {\cal N}\left( {0,\sigma _{{V_i} \to {V_n}}^2} \right)$ are the ToA measurement errors caused by UAVs' internal noise and jamming; ${\tau _f}$ denotes the time-of-flight (ToF) between these two UAVs.

Then, the estimated ToF is given by
\begin{equation}\tag{64}
\begin{split}
{{\hat \tau }_f} = \frac{1}{2}\left(\! {{{\hat \tau }_1} - {{\hat \tau }_2}} \!\right) &= {\tau _f} \!\cdot\! \left(\! {1 \!+\! {\delta _{{V_n}}}} \!\right) + \frac{{{e_{{V_n} \to {V_i}}}}}{c} \!\cdot\! \frac{{\left(\! {1 \!+\! {\delta _{{V_i}}}} \!\right)}}{{2\left(\! {1 \!+\! {\delta _{{V_n}}}} \!\right)}}\\
&+ \frac{{{e_{{V_i} \to {V_n},1}}}}{c} - \frac{{{e_{{V_i} \to {V_n},2}}}}{{2c}}.
\end{split}
\end{equation}

Since the values of ${\delta _{{V_n}}}$ and ${\delta _{{V_i}}}$ are commonly very small in practice, the above equation can be rewritten as
\begin{equation}\tag{65}
{\tau _f} \approx {\tau _f} + \frac{{{e_{{V_n} \to {V_i}}}}}{{2c}} + \frac{{{e_{{V_i} \to {V_n},1}}}}{c} - \frac{{{e_{{V_i} \to {V_n},2}}}}{{2c}}.
\end{equation}

Therefore, the range measurement obtained by DR-TWR technique can be expressed as
\begin{equation}\tag{66}
\begin{split}
{r_{{V_n} \!\to\! {V_i}}} \!=\! c \!\cdot\! {{\hat \tau }_f} &= c \!\cdot\! {\tau _f} + \frac{{{e_{{V_n} \!\to\! {V_i}}}}}{2} + {e_{{V_i} \!\to\! {V_n},1}} - \frac{{{e_{{V_i} \!\to\! {V_n},2}}}}{2}\\
&= r_{{V_n} \to {V_i}}^ *  + {n_{{V_n} \to {V_i}}},
\end{split}
\end{equation}
where ${n_{{V_n} \!\to\! {V_i}}} \!=\! \frac{{{e_{{V_n} \!\to\! {V_i}}}}}{2} \!+\! {e_{{V_i} \!\to\! {V_n},1}} \!-\! \frac{{{e_{{V_i} \!\to\! {V_n},2}}}}{2}$. Since the noise terms ${e_{{V_n} \to {V_i}}}$, ${e_{{V_i} \to {V_n},1}}$ and ${e_{{V_i} \to {V_n},2}}$ are independent of each other, the range measurement error ${n_{{V_n} \to {V_i}}}$ follows a zero-mean Gaussian distribution: ${n_{{V_n} \to {V_i}}} \sim {\cal N}\left( {0,\frac{1}{4}\sigma _{{V_n} \to {V_i}}^2 + \frac{5}{4}\sigma _{{V_i} \to {V_n}}^2} \right)$.

\bibliographystyle{IEEEtran}
\bibliography{mybib}

\end{document}